\documentclass[10pt,journal]{IEEEtran}
\usepackage{graphicx}
\usepackage{orcidlink}
\usepackage{float}
\usepackage{bm}
\usepackage{subfig}
\usepackage{amsmath,amsfonts}
\usepackage{mathtools}
\usepackage{algorithm}
\usepackage{algorithmic}
\usepackage[normalem]{ulem}
\usepackage[tableposition=top]{caption}
\usepackage{url}
\usepackage{booktabs}

\usepackage{array}
\usepackage{cite}
\usepackage{color}
\usepackage{cleveref}
\usepackage{geometry}
\usepackage{amssymb}
\usepackage{booktabs}
\usepackage{array}
\usepackage{pifont}
\usepackage{cuted}
\usepackage{lipsum} 
\usepackage{multicol}
\newgeometry{left=1.4cm,right=1.4cm,bottom=1.45cm,top=1.2cm}

\begin{document}

\title{Blind OFDM-ISAC Relying on Asymmetric Modem Constellations}
\author{
Henglin Pu, Ahmad Musallam, Husheng Li, \IEEEmembership{Senior Member, IEEE}, and Lajos Hanzo, \IEEEmembership{Life Fellow, IEEE} 


\thanks{Henglin Pu and Husheng Li are with the Elmore Family School of Electrical and Computer Engineering, Purdue University, West Lafayette, Indiana, USA (e-mail:
 pu36@purdue.edu, husheng@purdue.edu).}%
\and
\thanks{Ahmad Musallam and Husheng Li are with the School of Aeronautics and
 Astronautics, Purdue University, West Lafayette, Indiana, USA (e-mail: amusalla@purdue.edu, husheng@purdue.edu).}%
 \and
 \thanks{L. Hanzo is with the School of Electronics and Computer Science, University of Southampton, Southampton SO17 1BJ, U.K (e-mail:
lh@ecs.soton.ac.uk).}
}

\maketitle

\thispagestyle{empty}

\IEEEpeerreviewmaketitle

\begin{abstract}
Integrated sensing and communication (ISAC) is increasingly expected to operate under aggressive spectrum reuse, where co-channel orthogonal frequency division multiplexing (OFDM) interference can be catastrophic for data recovery on the time-frequency (TF) grid. We show that supporting blind ISAC is feasible by exploiting a fundamental asymmetry in the impact of co-channel OFDM interference: while communication is fragile on the TF grid, sensing depends on structured physical parameters whose signatures remain identifiable by relying on higher-order statistics. Based on this observation, we construct a fourth-order measurement tensor from the received OFDM signal whose coherent component preserves the delay-, Doppler-, and angle-dependent phase evolution of each source. We then develop a three-dimensional higher-order-statistics (HOS) based periodogram for iterative peak search and refinement to jointly estimate both range, velocity, and angle in the presence of unknown co-channel interferers. We further exploit constellation asymmetry to resolve the remaining phase ambiguities of blind recovery, enabling blind coherent demodulation via minimum constellation fitting. We also benchmark the performance through matched data-aided and stochastic Cramér–Rao lower bounds. We then quantify the cost of signal blindness. Simulations and experimental validations demonstrate reliable radar parameter estimation together with effective communication demodulation even when the TF-domain link is severely interfered with.
\end{abstract}


\section{Introduction}
\label{sec:introduction}

Integrated sensing and communication (ISAC) has emerged as a cornerstone technology for next-generation wireless networks, where the same infrastructure is expected to exchange information and simultaneously sense the surrounding physical environment~\cite{ISAC1, ISAC2}. By sharing spectrum, hardware, and signal processing resources between radar sensing and wireless communications, ISAC improves spectral efficiency~\cite{pu2025wideband} and supports a wide range of applications~\cite{ISAC_applications}. Orthogonal frequency division multiplexing~(OFDM), which has been widely adopted in modern wireless standards, is particularly attractive for ISAC because of its flexible bandwidth allocation, low-complexity single-tap equalization, and compatibility with existing communication infrastructures.

Given the increasing demand for fine-grained sensing, ISAC is evolving from isolated point-to-point sensing toward dense network-level sensing architectures, where multiple base stations, users, and ISAC nodes may participate in sensing over shared-spectral resources~\cite{11479627}. However, this dense deployment also introduces a fundamental challenge, namely co-channel interference~\cite{ISAC_interfrence}. In uncoordinated environments, different users or ISAC nodes may transmit independent data-bearing OFDM signals over the same time-frequency~(TF) resources. Consequently, an ISAC receiver may have to detect weak target echoes, while simultaneously receiving strong co-channel communication signals from unintended illuminators or neighboring base stations~(BS)~\cite{ISAC_SP}. This near-far power disparity may severely degrade the sensing dynamic range, especially when the receiver has no access to the transmitted data symbols, pilots, or clean reference channels of the interfering sources.

\begin{figure*}[!t]
 \centering
 \includegraphics[width=5.3in]{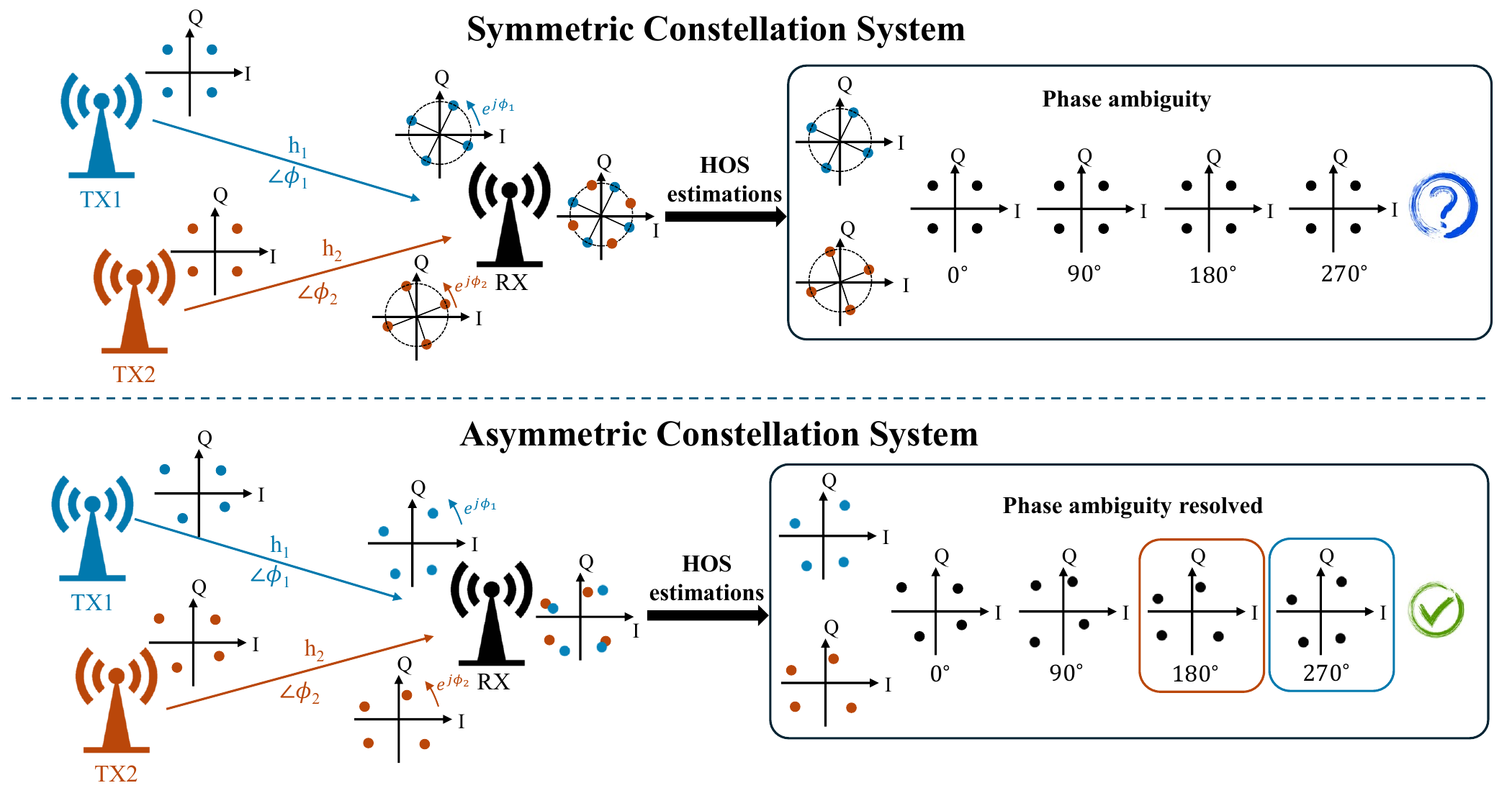}\\
 \caption{Illustration of the symmetric and asymmetric constellation diagrams.}
 \label{fig:APSK-16 diagram}
\end{figure*}

A key observation motivating this paper is that co-channel interference does not affect communication and sensing in the same manner. Communication reliability depends on recovering the transmitted symbol values on the TF grid, where independently generated OFDM signals collide directly. By contrast, sensing depends on estimating a small number of physical parameters, such as delay, Doppler, and angle, which induce structured phase progressions across frequency, slow time, and space. Hence, a pair of signals that severely interfere in the TF domain may remain distinguishable in the physical sensing domain, provided that their underlying delay-Doppler-angle supports are different. This motivates the development of a blind OFDM-ISAC receiver that can exploit the hidden physical structure of co-channel data-bearing OFDM mixtures without relying on pilots or reference waveforms.

\subsection{Review of Existing Works}


The problem of interference management in ISAC has attracted significant recent attention. Existing solutions often rely on coordinated waveform-, beamforming-, or precoding-designs to suppress harmful interference or exploit constructive interference~\cite{ZHANG2023103861,Wang2025_SLP_ISAC,Liao2023_SLP_ISAC,Chen2023_SLP_MIMO_ISAC,Cai2024_SLP_SIC_ISAC}. In parallel, OTFS-based ISAC waveforms have been investigated in high-mobility scenarios, since the delay-Doppler representation naturally supports target sensing~\cite{Shi2024_OTFS_ISAC,Wei2025_OTFS_V2X_ISAC}. Despite their effectiveness, these approaches typically require accurate CSI, tight synchronization, pilot signaling, or direct control over the transmitted waveforms, which are difficult to guarantee in passive or uncoordinated ISAC scenarios involving independent data-bearing transmitters.

Blind estimation methods provide a natural alternative, when pilots or reference waveforms are unavailable. In the communications literature, cyclostationarity has been widely exploited for blind OFDM signal recognition and parameter estimation, for example through cyclic-correlation or joint delay--cyclic-frequency features~\cite{sun2014cyclostationarity}. Higher-order statistics (HOS) have also been used for blind OFDM modulation classification, where fourth-order cumulants are extracted from the received OFDM signals to identify modulation formats without prior knowledge of the signal parameters or channel statistics~\cite{Gupta2020_BlindOFDMModClass, 11432914}. In addition to recognition and classification, classical blind OFDM channel estimation and equalization methods exploit communication-domain structures such as cyclic prefices, virtual carriers, second-order statistics, or transmitter-side precoding to estimate the channel without explicit pilots~\cite{muquet2002subspace,bolcskei2002blind}. While these treatises demonstrate that OFDM signals contain useful pilot-free statistical and structural information, their objectives remain centered on waveform recognition, modulation classification, or channel identification. They do not directly estimate the physical delay, Doppler, and angle parameters of multiple co-channel OFDM sources observed by an ISAC receiver, nor do they address blind sensing-parameter estimation and symbol recovery under co-channel OFDM-ISAC interference.

In the ISAC literature, dual-blind deconvolution (DBD)~\cite{DBD} has been proposed for overlaid radar and communication signals with unknown transmitted waveforms and propagation channels. By exploiting continuous-domain sparsity, DBD and its multi-antenna extensions~\cite{9746868,Jacome2024_MDBD} can recover both radar and communication parameters via atomic-norm-based optimization over delay, Doppler, and direction-of-arrival structures. However, these methods mainly target heterogeneous radar-communication coexistence, where the radar and communication components occupy distinct waveform structures or signal subspaces, rather than the co-channel OFDM-ISAC setting, where multiple independent sources share the same OFDM waveform family and collide on the same TF grid. Recent pilot-free ISAC methods based on structured matrix factorization~\cite{yuan2024integrated} or lifted atomic-norm minimization~\cite{Valiulahi2026_LANM_ISAC} have demonstrated blind joint sensing and symbol recovery, but they either rely on differential modulation to bypass phase ambiguity or impose low-dimensional transmit subspace models, and do not address ambiguity-resolved coherent demodulation under co-channel OFDM collisions.

The distinctive features of the proposed framework are boldly contrasted to representative blind OFDM, blind radar-communications, and pilot-free ISAC studies in Table~\ref{tab:review_existing_works}. As seen from the table, existing contributions address important subsets of the problem, but none of them simultaneously considers blind co-channel OFDM sensing, blind sensing-parameter estimation, communication-symbol recovery, and ambiguity-resolved blind demodulation. The proposed framework is further supported by Cram\'er-Rao lower bound (CRLB) analysis and verified by hardware validation.

\begin{table*}[t]
\centering
\captionsetup{font=footnotesize,labelfont=footnotesize}
\caption{Contrasting Our Contribution to the Existing Literature}
\label{tab:review_existing_works}
\begin{tabular}{|l|c|c|c|c|c|c|c|c|c|}
\hline

& \cite{sun2014cyclostationarity}
& \cite{Gupta2020_BlindOFDMModClass}
& \cite{11432914}
& \cite{muquet2002subspace}
& \cite{bolcskei2002blind}
& \cite{DBD}
& \cite{yuan2024integrated}
& \cite{Valiulahi2026_LANM_ISAC}
& \textbf{Proposed} \\
\hline
ISAC system
&  &  &  $\checkmark$
&  & 
& $\checkmark$  & $\checkmark$ & $\checkmark$
& $\checkmark$ \\
\hline
Co-channel signal model
&  &   & $\checkmark$
&  & 
& $\checkmark$  & $\checkmark$ & 
& $\checkmark$ \\
\hline
Shared sensing/communication waveform
&  &   & $\checkmark$
&  & 
&   & $\checkmark$ & $\checkmark$
& $\checkmark$ \\
\hline
Pilot/reference-free receiver
& $\checkmark$ & $\checkmark$  & $\checkmark$
& $\checkmark$ & $\checkmark$
& $\checkmark$ & $\checkmark$ & $\checkmark$
& $\checkmark$ \\
\hline
Blind sensing-parameter estimation
&  &   & $\checkmark$
&  & 
& $\checkmark$ & $\checkmark$ & $\checkmark$
& $\checkmark$ \\
\hline
Higher-order-statistics-based estimation
&  & $\checkmark$  & $\checkmark$
&  & 
&   &  & 
& $\checkmark$ \\
\hline
Blind communication-symbol recovery
&  &   & 
&  & $\checkmark$
& $\checkmark$  & $\checkmark$ & $\checkmark$
& $\checkmark$ \\
\hline
Ambiguity-resolved blind demodulation
&  &   & 
&  & 
& $\checkmark$  &  & $\checkmark$
& $\checkmark$ \\
\hline
CRLB / performance bounds
&  &   & 
&  & 
& $\checkmark$  &  & $\checkmark$
& $\checkmark$ \\
\hline
Experimental / hardware validation
&  & $\checkmark$  & $\checkmark$
&  & 
&   &  & 
& $\checkmark$ \\
\hline
\textbf{Asymmetric constellation design}
&  &   & 
&  & 
&   &  & 
& $\checkmark$ \\
\hline
\end{tabular}
\end{table*}

\subsection{Contributions of the Present Work}

Motivated by the above knowledge-gaps, this paper develops a blind space-time-frequency OFDM-ISAC framework that exploits HOS for blind sensing and communication recovery. The proposed framework is based on two forms of asymmetry. The first is the asymmetry between TF-domain communication interference and physical-domain sensing structure. While independent OFDM sources collide on the TF grid, their delay-, Doppler- and angle-dependent phase progressions remain structured across frequency, slow time, and space. The second is constellation asymmetry. As illustrated in Fig.~\ref{fig:APSK-16 diagram}, symmetric constellations such as standard 16QAM retain rotational symmetry, so blind recovery leaves residual phase ambiguities because multiple rotations produce equivalent constellation patterns. By contrast, an asymmetric constellation breaks this rotational symmetry and provides an intrinsic phase reference for ambiguity resolution.

More specifically, we show that although the second-order statistics of independent co-channel OFDM sources are largely uninformative for physical separation, their fourth-order statistics preserve deterministic phase evolution induced by delay, Doppler, and angle. By arranging these fourth-order measurements into a three-dimensional tensor over the frequency, time, and space dimensions, the receiver can blindly estimate the physical sensing parameters without pilots, reference channels, or without demodulating the interfering data. The resultant blind parameter estimates then provide sufficiently accurate CSI for multi-source separation and communication-symbol recovery. However, the blind processing pipeline still leaves residual ambiguities, namely a per-stream complex factor and a potentially possible stream permutation. These ambiguities are largely harmless for sensing, but critical for coherent demodulation. By adopting an asymmetric constellation and applying minimum constellation fitting to the post-equalization streams, the proposed receiver resolves these ambiguities and facilitates blind coherent demodulation. The main contributions of this paper are summarized as follows:
\begin{itemize}
    \item
    We reveal and exploit the asymmetric impact of co-channel interference on both communication and sensing in OFDM-ISAC systems. Specifically, while independent OFDM sources collide on the TF grid for communication-symbol recovery, their physical sensing signatures remain structured across the frequency, slow time, and space dimensions, enabling reliable sensing even under low signal-to-interference-ratio (SIR) conditions.

    \item 
    We construct a fourth-order space-time-frequency measurement tensor whose coherent component preserves the \mbox{delay-,} Doppler-, and angle-dependent phase progressions of the sources. Based on this tensor, we develop a 3D HOS periodogram associated with iterative peak enumeration and refinement for joint range, velocity, and angle estimation without pilots or reference channels.

    \item 
    We propose an ambiguity-resolved blind coherent demodulation framework. After HOS-based sensing and spatial separation, asymmetric constellations and minimum constellation fitting are used for resolving the residual rotation, scale, and permutation ambiguities, thereby enabling coherent symbol recovery without pilots or differential coding.

    \item 
    We derive matched data-aided and stochastic CRLBs for the 3D sensing problem, which quantify the fundamental cost of blindness. The resultant bounds are used for characterizing the implementation loss imposed by fourth-order processing and for benchmarking the proposed estimator in both simulations and hardware experiments.
\end{itemize}

\subsection{Organization of the Paper}

The remainder of this paper is organized as follows. The system model is introduced in Section~\ref{sec:system_model}. The proposed HOS-based blind sensing method is presented in Section~\ref{sec:proposed_method}. Blind data demodulation is developed in Section~\ref{sec:data_recovery}, followed by the asymmetric-constellation design in Section~\ref{sec:asym_constellation}. The CRLB analysis is presented in Section~\ref{sec:CRLB}. Numerical and experimental results are provided in Section~\ref{sec:results}. Finally, conclusions are drawn in Section~\ref{sec:conclusion}.

\section{System Model}
\label{sec:system_model}

Consider a generic OFDM-based ISAC system, where the receiver is equipped with a uniform linear array (ULA) of $M$ antennas, but our formulation extends straightforwardly to 2D arrays as well. The receiver observes a superposition of $P$ independent sources sharing the same frequency band. Let $p=0$ denote the desired target echo (sensing signal) and $p \ge 1$ denote co-channel interference sources (e.g., communication users or other radar transmitters) sharing the same frequency band.

\subsection{Transmitted Signal Model}
The OFDM-based ISAC system operates over a total bandwidth $B$ at carrier frequency $f_c$, which is partitioned into $K$ orthogonal subcarriers with frequency spacing $\Delta f = B/K$. In the time domain~(TD), transmission is organized into $N_{\mathrm{sym}}$ consecutive OFDM symbols. The duration of the OFDM symbol interval is given by $T_{u} = 1/\Delta f$. To mitigate inter-symbol interference (ISI), a cyclic prefix (CP) of length $T_{\mathrm{cp}}$ is appended to each symbol, resulting in a total OFDM symbol duration of $T_{sym} = T_u + T_{cp}$.

The complex baseband signal transmitted by the $p$-th source during the $n$-th OFDM symbol period is given by
\begin{small}
\begin{equation}
    s_p(t) = \sum_{n=0}^{N_{sym}-1} \sum_{k=0}^{K-1} X_{p,k}[n] e^{j2\pi k \Delta f (t - n T_{sym})} \text{rect}\left(\frac{t - n T_{sym}}{T_{sym}}\right),
\end{equation}
\end{small}
where $X_{p,k}[n]$ represents the complex data symbol mapped to the $k$-th subcarrier of the $n$-th symbol and $\mathrm{rect}(\cdot)$ denotes the standard rectangular pulse. 


\subsection{Channel Geometry and Parameters}
The physical propagation associated with each source is described in the DD domain. For the $p$-th source, we collect its continuous channel parameters into the vector
\begin{equation}\label{eq:channel_parameters}
    \boldsymbol{\xi}_p \triangleq [\tau_p, \nu_p, \theta_p]^{\mathsf T},
\end{equation}
where $\tau_p$, $\nu_p$, and $\theta_p$ denote the delay, Doppler frequency shift, and angle of arrival (AoA) of the $p$-th source at the receiver array, respectively.


The delay $\tau_p \in \mathbb{R}_+$ represents the propagation delay associated with the $p$-th path. More generally, $\tau_p$ denotes the effective propagation delay of that source-path pair. When a geometric interpretation is needed, it can be related to the propagation distance through the speed of light $c$.

The Doppler frequency $\nu_p \in \mathbb{R}$ characterizes the frequency shift imposed by the relative radial motion between the receiver and the $p$-th source. It governs the phase evolution across OFDM symbols through factors of the form $e^{j 2\pi n T_{\mathrm{sym}} \nu_p}$, and is proportional to the radial velocity $v_p$. 

The AoA $\theta_p \in [-\pi/2,\pi/2)$ denotes the direction of the impinging plane wave measured relative to the array broadside. For a ULA with inter-element spacing $d$, the corresponding array steering vector is given by
\begin{equation}
\mathbf{a}(\theta_p)
=
\begin{bmatrix}
1 &
e^{-j 2\pi \frac{d}{\lambda}\sin(\theta_p)} &
\cdots &
e^{-j 2\pi \frac{d}{\lambda}(M-1)\sin(\theta_p)}
\end{bmatrix}^{\mathsf T}.
\end{equation}
Hence the spatial response at antenna index $m$ is $a_m(\theta_p)=e^{-j 2\pi \frac{d}{\lambda} m \sin(\theta_p)}$ for $m=0,\ldots,M-1$. The triplet $(\tau_p,\nu_p,\theta_p)$ therefore specifies the location and motion of the $p$-th source in the delay--Doppler domain.

\subsection{Received Signal Model}
After down-conversion to the base band, CP removal, and discrete Fourier transform (DFT), the received signal in the frequency domain~(FD) at the $m$-th antenna, $k$-th subcarrier, and $n$-th OFDM symbol can be expressed as a superposition of the channel responses scaled by the transmitted data. Under the assumption that the signal bandwidth is much lower than the carrier frequency, the array steering vector is approximately frequency-flat.
Moreover, we assume a limited-Doppler regime in which the channel can be treated as time-invariant over the useful OFDM symbol duration. Equivalently, the normalized Doppler $\frac{v_p f_c}{c}$ is sufficiently low, so that the Doppler-induced inter-carrier interference (ICI) is negligible and the Doppler shift manifests itself primarily as a symbol-to-symbol phase progression. Let $Y_{k,n,m}$ denote the received sample. With these assumptions, the received signal is modeled as:
\begin{equation}
    Y_{k,n,m} = \sum_{p=0}^{P-1} \alpha_p \cdot X_{p,k}[n] \cdot H(k,n,m; \bm{\xi}_p) + W_{k,n,m},
    \label{eq:rx_signal}
\end{equation}
where $\alpha_p$ is the complex channel gain (including path loss and radar cross-section), $W_{k,n,m} \sim \mathcal{CN}(0, \sigma_n^2)$ is the additive white Gaussian noise and $H(k,n,m; \bm{\xi}_p)$ is the 3D channel response function, factorized as:
\begin{align}
    H(k,n,m; \bm{\xi}_p) &= \underbrace{e^{-j2\pi k \Delta f \tau_p}}_{\text{Frequency Domain}} \cdot \underbrace{e^{j2\pi n T_{sym} \nu_p}}_{\text{Time Domain}} \cdot \underbrace{e^{-j 2\pi \frac{d}{\lambda} m \sin(\theta_p)}}_{\text{Spatial Domain}}.
\end{align}

\subsection{Problem Formulation}

The ISAC receiver collects the received samples across FD (subcarrier index), TD (OFDM symbol index), and space (antenna index) into the third-order observation tensor $\mathcal{Y} \in \mathbb{C}^{K \times N_{\mathrm{sym}} \times M}$, whose $(k,n,m)$-th entry $Y[k,n,m]$ denotes the signal observed on subcarrier $k \in \{0,\ldots,K-1\}$, OFDM symbol index $n \in \{0,\ldots,N_{\mathrm{sym}}-1\}$, and antenna element $m \in \{0,\ldots,M-1\}$.

For the $p$-th source, based on the continuous channel parameters in \eqref{eq:channel_parameters}, we define the aggregate set of unknown sensing parameters as
\begin{equation}
    \boldsymbol{\Xi} \triangleq \{\boldsymbol{\xi}_0, \ldots, \boldsymbol{\xi}_{P-1}\}.
\end{equation}
The corresponding transmitted data symbols are stacked into the tensor as $\mathcal{X} \in \mathbb{C}^{K \times N_{\mathrm{sym}} \times P}$,
with entries $X[k,n,p] = X_{p,k}[n]$, where $X_{p,k}[n]$ denotes the symbol of the $p$-th source on subcarrier $k$ during the OFDM symbol index $n$. 


The blind sensing problem considered in this work is to estimate the collection of channel parameters
$\boldsymbol{\Xi}\triangleq\{\boldsymbol{\xi}_p\}_{p=0}^{P-1}$ from the received observation tensor $\mathcal{Y}$
without access to the transmitted data tensor $\mathcal{X}$. In particular, the symbols $X_{p,k}[n]$ are unobserved and
act as random nuisance variables. The objective is to construct a set of estimators:
\begin{equation}
    \hat{\boldsymbol{\xi}}_p = f_p(\mathcal{Y}), \quad p=0,\ldots,P-1,
\end{equation}
so that the estimation error is minimized in the mean-squared sense. Specifically, we consider the mean-squared error (MSE)
\begin{equation}
    \mathrm{MSE}(\hat{\boldsymbol{\xi}}_p)
    \triangleq \mathbb{E}\!\left[\left\|\hat{\boldsymbol{\xi}}_p-\boldsymbol{\xi}_p\right\|^2\right],
\end{equation}
where the expectation is taken with respect to the unknown data symbols and the receiver noise. Throughout this paper, the data symbols
are assumed to be i.i.d. and mutually independent across different sources.

\section{Joint Space-Delay-Doppler Estimation via HOS Periodogram}
\label{sec:proposed_method}


To recover the sensing parameters $\bm{\Xi}=\{\tau_p,\nu_p,\theta_p\}_{p=0}^{P-1}$, we assume that the receiver knows the OFDM waveform parameters, including the carrier frequency, subcarrier spacing, and constellation family, but does not know the transmitted symbol realizations $X_{p,k}[n]$. These symbol values are treated as random nuisance variables. Based on this model, we develop a constructive estimator using the fourth-order statistics of the received signal. Unlike subspace-based methods such as ESPRIT, which can be sensitive to model-order mismatch and cross-terms~\cite{subspace_problem1, subspace_problem2}, we develop a 3D HOS periodogram with separable parabolic interpolation for robust parameter estimation. This approach converts the blind sensing problem into a three-dimensional spectral peak search in a virtual fourth-order domain, thereby exploiting the structured phase evolution of the received signal to achieve high-resolution estimation.



\subsection{Fourth-Order Virtual Tensor Construction}
Let $\mathcal{Y} \in \mathbb{C}^{K \times N_{sym} \times M}$ denote the received signal tensor. We construct the fourth-order measurement tensor $\mathcal{Z}$ via the element-wise fourth power operation $\mathcal{Z}_{k,n,m} \triangleq (Y_{k,n,m})^4$. Upon expanding $\mathcal{Z}_{k,n,m}$ using \eqref{eq:rx_signal}, we have 
\begin{equation}
Z_{k,n,m} = \left(\sum_{p=0}^{P-1} S_{p,k,n,m}+W_{k,n,m}\right)^4,
\label{eq:Z_expand}
\end{equation}
where $S_{p,k,n,m}\triangleq \alpha_p X_{p,k}[n] H(k,n,m; \bm{\xi}_p)$. Under the assumptions that the source symbols are mutually independent, independent of the receiver noise, and satisfy $E[X_{p,k}[n]]=0$ and $E[X_{p,k}[n]^2]=0$. The odd mixed terms vanish in expectation. Moreover, for proper complex AWGN, we have $E[W_{k,n,m}^4]=0$. Therefore, the coherent contribution is
\begin{equation}
E\!\left[S_{p,k,n,m}^4\right]
=
\alpha_p^4 \mu_4 H(k,n,m;\xi_p)^4,
\qquad
\mu_4 \triangleq E[X^4].
\label{eq:self_term_expectation}
\end{equation}
It follows then that
\begin{equation}
E[Z_{k,n,m}]=\sum_{p=0}^{P-1} \alpha_p^4 \mu_4 H(k,n,m; \bm{\xi}_p)^4
+ \varepsilon_{k,n,m},
\label{eq:EZ_decomp}
\end{equation}
where $\varepsilon_{k,n,m}$ collects the remaining even-order mixed terms. Under the idealized model, these terms vanish in expectation; for finite data, however, they act as residual self-noise and inter-source clutter. According to \eqref{eq:EZ_decomp}, the expectation of the fourth-order tensor contains a coherent sum of separable 3D exponentials, one from each source. In particular, the contribution of the $p$-th source is given by
\begin{equation}
    \begin{aligned}
\mathbb{E}\!\left[\mathcal{Z}_{k,n,m}^{(p)}\right]
&= \tilde{\alpha}_p\,
\underbrace{e^{-j 2\pi k (4\Delta f \tau_p)}}_{\text{Frequency domain}}\, \underbrace{e^{j 2\pi n (4 T_{\mathrm{sym}} \nu_p)}}_{\text{Time domain}}\, \\
& \qquad \qquad \qquad \qquad \qquad 
\underbrace{e^{-j 2\pi m \left(4\frac{d}{\lambda}\sin(\theta_p)\right)}}_{\text{Spatial domain}},
\end{aligned}\label{eq:virtual_signal}
\end{equation}
where $\tilde{\alpha}_p \triangleq \alpha_p^4 \mu_4$. Equation \eqref{eq:virtual_signal} implies that the fourth-order transformation maps the physical parameters to a virtual parameter space having scaled resolutions of:
\begin{itemize}
    \item Virtual Delay Frequency: $\tilde{f}_{\tau} = 4 \Delta f \tau_p$.
    \item Virtual Doppler Frequency: $\tilde{f}_{\nu} = 4 T_{sym} \nu_p$.
    \item Virtual Spatial Frequency: $\tilde{f}_{\theta} = 4\frac{d}{\lambda}\sin(\theta_p).$
\end{itemize}



Note that a fourth-order cumulant could further suppress Gaussian noise contributions, but it requires additional lower-order moment corrections, increasing complexity and finite-sample sensitivity. Since $\mathbb{E}[W^4]=0$ for proper complex AWGN, the proposed 
$Y^4$ statistics already preserve the desired coherent component, while remaining implementation-friendly.

\subsection{Super-Resolution Spectral Estimation}
To estimate the sensing parameters from the finite-sample tensor $\mathcal{Z}$, we interpret the problem as detecting multiple spectral peaks in the 3D HOS spectrum. Since the number of active sources is generally unknown a priori, we do not assume that $P$ is given. Instead, we adopt an iterative peak-detection procedure relying on adaptive stopping.

\subsubsection{3D Discrete Fourier Transform}
We compute the 3D DFT of the tensor $\mathcal{Z}$, using substantial zero-padding to better approximate the continuous-frequency spectrum and mitigate the “picket-fence” effect~\cite{1455106}. Let $\mathbf{N}_{\mathrm{fft}}=[N_\tau,N_\nu,N_\theta]$ denote the FFT dimensions, where $N_\tau \gg K$, $N_\nu \gg N_{\mathrm{sym}}$, and $N_\theta \gg M$. The resultant spectrum is
\begin{equation}
    \mathcal{S}[u,v,w]
    =
    \sum_{k=0}^{K-1}\sum_{n=0}^{N_{\mathrm{sym}}-1}\sum_{m=0}^{M-1}
    \mathcal{Z}_{k,n,m}
    e^{-j2\pi\left(\frac{uk}{N_\tau}+\frac{vn}{N_\nu}+\frac{wm}{N_\theta}\right)}.
    \label{eq:HOS_spectrum}
\end{equation}

\subsubsection{Iterative Spectral Enumeration}
From \eqref{eq:virtual_signal}, each source contributes a separable 3D complex exponential in the fourth-order domain. Therefore, \eqref{eq:HOS_spectrum} can be interpreted as the matched response of $\mathcal{Z}$ to candidate virtual delay-Doppler-angle atoms on the FFT grid. At iteration $p$, the dominant residual component is thus estimated by the peak of the residual spectrum:
\begin{equation}
    (u_0,v_0,w_0)\in
    \arg\max_{u,v,w}
    \left|\mathcal{S}^{(p)}[u,v,w]\right|.
    \label{eq:find_peak}
\end{equation}
Here, $\mathcal{S}^{(1)}=\mathcal{S}$ denotes the initial spectrum, while $\mathcal{S}^{(p)}$ is the residual spectrum at iteration $p$.

\begin{algorithm}[t]
\caption{Blind Joint Estimation via HOS Periodogram}
\label{alg:blind_hos}
\begin{algorithmic}[1]
\REQUIRE Received tensor $\mathcal{Y}\in\mathbb{C}^{K\times N_{\mathrm{sym}}\times M}$; fourth moment $\mu_4$; FFT sizes $(N_\tau,N_\nu,N_\theta)$; masking radius $R_{\mathrm{mask}}$; detection threshold $\gamma$.
\ENSURE Source number $\hat{P}$ and estimates $\{(\hat{\tau}_p,\hat{\nu}_p,\hat{\theta}_p,\hat{\alpha}_p)\}_{p=1}^{\hat{P}}$.

\STATE Form the fourth-order tensor $\mathcal{Z}$ from $\mathcal{Y}$ according to \eqref{eq:virtual_signal}.
\STATE Compute the 3D HOS spectrum $\mathcal{S}^{(1)}$ according to \eqref{eq:HOS_spectrum}.
\STATE Set $G_{\mathrm{fft}}=K N_{\mathrm{sym}} M$ and $p=1$.

\WHILE{true}
    \STATE Find the dominant residual peak $(u_0,v_0,w_0)$ using \eqref{eq:find_peak}.
    \STATE Estimate the local background power $\widehat{\sigma}^{2}_{\mathrm{loc}}$ around $(u_0,v_0,w_0)$.
    \IF{$\frac{|\mathcal{S}^{(p)}[u_0,v_0,w_0]|^2}{\widehat{\sigma}^{2}_{\mathrm{loc}}} \le \gamma$}
        \STATE Set $\hat{P}=p-1$ and \textbf{break}.
    \ENDIF
    \STATE Refine the peak location via \eqref{eq:parabolic} to obtain $(\hat{u},\hat{v},\hat{w})$.
    \STATE Map $(\hat{u},\hat{v},\hat{w})$ to $(\hat{\tau}_p,\hat{\nu}_p,\hat{\theta}_p)$ using \eqref{eq:inverse_mapping}.
    \STATE Estimate the complex channel gain $\hat{\alpha}_p$ using \eqref{eq:complex_gain_est}.
    \STATE Update the residual spectrum using \eqref{eq:spectral_masking}.
    \STATE Set $p\leftarrow p+1$.
\ENDWHILE
\end{algorithmic}
\end{algorithm}

To mitigate the finite-grid resolution imposed by the FFT, we refine each coordinate by separable parabolic interpolation. For one dimension $d\in\{u,v,w\}$, let $y_0$ denote the peak magnitude at the maximizer, and let $y_{-1}$ and $y_{+1}$ denote the magnitudes at its pair of adjacent bins along dimension $d$, with the other two coordinates fixed. Assuming the local peak shape to be well approximated by a quadratic function, the sub-bin offset is estimated from the vertex of the fitted parabola as
\begin{equation}
    \delta_d=
    \frac{0.5\,(y_{-1}-y_{+1})}
    {y_{-1}-2y_0+y_{+1}},
    \label{eq:parabolic}
\end{equation}
which yields the refined coordinates
$\hat{u}=u_0+\delta_u$,
$\hat{v}=v_0+\delta_v$, and
$\hat{w}=w_0+\delta_w$.

The refined virtual frequencies are then mapped back to the physical parameters by inverting the scaling in \eqref{eq:virtual_signal}:
\begin{equation}
\begin{aligned}
\hat{\tau}_p &= -\frac{\hat{u}}{4N_\tau\Delta f}, \qquad
\hat{\nu}_p = \frac{\hat{v}}{4N_\nu T_{\mathrm{sym}}},\\
\hat{\theta}_p &= \arcsin\!\left(-\frac{\hat{w}\lambda}{4N_\theta d}\right).
\end{aligned}
\label{eq:inverse_mapping}
\end{equation}
Since the number of sources is unknown, we employ a CFAR-like stopping rule~\cite{4102829}. Let $\mathcal{M}^{(p)}[u,v,w]=\left|\mathcal{S}^{(p)}[u,v,w]\right|^2$ denote the residual HOS power spectrum. Around the candidate peak $(u_0,v_0,w_0)$, we estimate a local background level $\widehat{\sigma}^{2}_{\mathrm{loc}}$ from neighboring training cells outside a guard region. The peak is only accepted if $\frac{\mathcal{M}^{(p)}[u_0,v_0,w_0]}
    {\widehat{\sigma}^{2}_{\mathrm{loc}}}
    > \gamma,$
where $\gamma$ is a prescribed detection threshold. Otherwise, the iteration terminates, and the number of detected sources is set to $\hat{P}=p-1$.

To avoid repeatedly selecting the same source or its sidelobes, we suppress a neighborhood around each accepted peak. Specifically, after the $p$-th detection, the residual spectrum is updated as
\begin{equation}
    \left|\mathcal{S}^{(p+1)}[u,v,w]\right|=
    \begin{cases}
        0, \ \text{if } \left\|[u,v,w]-[u_0,v_0,w_0]\right\|_2<R_{\mathrm{mask}},\\
        \left|\mathcal{S}^{(p)}[u,v,w]\right|, \& \text{otherwise},
    \end{cases}
    \label{eq:spectral_masking}
\end{equation}
where $R_{\mathrm{mask}}$ is the masking radius. The overall procedure is summarized in Algorithm~\ref{alg:blind_hos}.

\subsection{Complex Amplitude Estimation}
Once the geometric parameters $\hat{\bm{\xi}}_p = \{\hat{\tau}_p, \hat{\nu}_p, \hat{\theta}_p\}$ are recovered, we estimate the complex channel gain $\alpha_p$. Since the fourth-order transformation preserves the complex phase information up to a four-fold ambiguity (due to the $\alpha_p^4$ term), we leverage the complex value of the spectral peak found in the previous step.

Let $\mathcal{V}_{peak}^{(p)} = \mathcal{S}^{(p)}[\hat{u}, \hat{v}, \hat{w}]$ be the complex value of the interpolated spectral peak for the $p$-th source. The complex channel gain is estimated by inverting the fourth-power scaling:
\begin{equation}
    \hat{\alpha}_p = \left( \frac{\mathcal{V}_{peak}^{(p)}}{\mu_4 \cdot G_{\mathrm{fft}}} \right)^{1/4},
    \label{eq:complex_gain_est}
\end{equation}
where $G_{\mathrm{fft}} = K N_{sym} M$ is the coherent processing gain and $\mu_4 = \mathbb{E}[X^4]$ is the known kurtosis of the constellation (e.g., $\mu_4 = -1$ for unit-power QPSK). 

It is important to note that blind estimation using fourth-order statistics inherently suffers from a phase ambiguity of $k\pi/2$ (for QPSK/QAM) because the operation $X^4$ is invariant to phase rotations of $\pi/2$. Consequently, the estimator recovers the true complex gain up to a quadrant rotation: $\hat{\alpha}_p = \alpha_p \cdot e^{j k \pi/2}$. This ambiguity does not affect the magnitude estimation $|\hat{\alpha}_p|$, which is critical for radar cross-section (RCS) analysis, nor does it affect the relative spatial phase which is captured by the spatial frequency $\hat{\theta}_p$.

\subsection{Complexity Analysis}
The computational complexity is dominated by the 3D FFT operation, scaling as $\mathcal{O}(\mathbf{N}_{\mathrm{fft}} \log \mathbf{N}_{\mathrm{fft}})$. While higher than 1D processing, these operations are highly parallelizable on modern hardware (FPGA/GPU). Crucially, the complexity is independent of the constellation order or the specific data sequences. This is in contrast to maximum likelihood (ML)  approaches which require exhaustive search over the codebook.

\section{Blind Data Demodulation}
\label{sec:data_recovery}

Once the channel parameters $\hat{\bm{\Xi}} = \{\hat{\tau}_p, \hat{\nu}_p, \hat{\theta}_p\}_{p=0}^{P-1}$ and complex amplitudes $\{\hat{\alpha}_p\}_{p=0}^{P-1}$ have been estimated, the ISAC receiver possesses full CSI for all sources. This allows the receiver to switch roles from a sensing device to a multi-user communication receiver, recovering the unknown information symbols $X_{p,k}[n]$ by inverting the estimated channel response. Note that after blind channel estimation, the remaining symbol-recovery stage becomes similar to a standard multi-user detector, as in non-orthogonal multiple access (NOMA) receivers~\cite{NOMA}.

\subsection{Instantaneous Mixing-Matrix Construction}
Since the sources occupy distinct locations in the DD domain, their effective array responses exhibit different phase evolutions across time and frequency. For each time--frequency resource element $(k,n)$, we collect the array outputs into a spatial snapshot vector
\begin{equation}
    \mathbf{r}[k,n] \triangleq \big[Y_{k,n,0},\, Y_{k,n,1},\, \ldots,\, Y_{k,n,M-1}\big]^{\mathsf T}\in\mathbb{C}^{M}.
\end{equation}

Using the estimated parameters $\{(\hat{\tau}_p,\hat{\nu}_p,\hat{\theta}_p,\hat{\alpha}_p)\}_{p=1,...,P}$, we construct an instantaneous mixing matrix $\hat{\mathbf{A}}[k,n]\in\mathbb{C}^{M\times P}$ whose $p$-th column represents the effective channel vector of source $p$ at $(k,n)$. Specifically, we define:
\begin{equation}
    \hat{\mathbf{a}}_p[k,n]
    \triangleq
    \hat{\alpha}_p\,
    \Phi_{\mathrm{TF}}\!\big(k,n;\hat{\tau}_p,\hat{\nu}_p\big)\,
    \mathbf{a}_s(\hat{\theta}_p),
    \label{eq:Mixing_a}
\end{equation}
where $\mathbf{a}_s(\hat{\theta}_p)$ denotes the array steering vector and
\begin{equation}
    \Phi_{\mathrm{TF}}\!\big(k,n;\hat{\tau}_p,\hat{\nu}_p\big)
    \triangleq
    e^{-j 2\pi k\Delta f\,\hat{\tau}_p}\,
    e^{j 2\pi nT_{\mathrm{sym}}\,\hat{\nu}_p}
\end{equation}
denotes the delay- and Doppler-induced phase progression on the TF grid. Stacking $\{\hat{\mathbf{a}}_p[k,n]\}_{p=0}^{P-1}$ yields
\begin{equation}
    \hat{\mathbf{A}}[k,n] \triangleq \big[\hat{\mathbf{a}}_0[k,n],\,\ldots,\,\hat{\mathbf{a}}_{P-1}[k,n]\big].
    \label{eq:Mixing_A}
\end{equation}

\subsection{Zero-Forcing Equalization}
At each TF grid point, the spatial observation is modeled as
\begin{equation}
    \mathbf{r}[k,n] \approx \hat{\mathbf{A}}[k,n]\mathbf{x}[k,n] + \mathbf{w}[k,n],
\end{equation}
where $\mathbf{x}[k,n]\triangleq [X_{0,k}[n],\ldots,X_{P-1,k}[n]]^{\mathsf T}$ collects the symbols from all sources. When $M\ge P$ and the columns of $\hat{\mathbf{A}}[k,n]$ are linearly independent with a high probability, the symbol vector can be recovered using a linear zero-forcing (ZF) equalizer formulated as
\begin{equation}
    \hat{\mathbf{x}}[k,n]
    =
    \big(\hat{\mathbf{A}}[k,n]^{H}\hat{\mathbf{A}}[k,n]\big)^{-1}
    \hat{\mathbf{A}}[k,n]^{H}\mathbf{r}[k,n].
    \label{eq:ZF}
\end{equation}
This operation forms $P$ spatial beams, providing unity response along each estimated direction $\hat{\theta}_p$, while suppressing the remaining sources through spatial nulling.


For the blind HOS-based estimator, the recovered symbol streams are only identifiable up to a permutation and a constant complex ambiguity. Specifically, since the blind sensing stage does not preserve source labels, the estimated mixing matrix can be expressed as
\begin{equation}
    \hat{\mathbf{A}}[k,n] \approx \mathbf{A}[k,n]\Pi D,
\end{equation}
where \(A[k,n]\) is the true mixing matrix, \(\Pi\) is an unknown permutation matrix, and \(D=\mathrm{diag}(\beta_0,\ldots,\beta_{P-1})\) collects unknown constant complex factors. Moreover, since the complex peak value in \eqref{eq:complex_gain_est} is proportional to \(\alpha_p^4\), the gain \(\alpha_p\) can only be recovered up to a fourth root of unity, i.e., we have:
\begin{equation}
  \hat{\alpha}_p=\alpha_p e^{j\frac{\pi}{2}\ell_p},\qquad \ell_p\in\{0,1,2,3\},  
\end{equation}
which results in a constant quadrant rotation in the recovered constellation. Thus, again, the blind HOS estimation determines each stream only up to an unknown permutation and a constant rotation/scale factor.

The residual ambiguities may be resolved by external references, such as pilots or differential modulation. However, these approaches either reduce spectral efficiency or sacrifice coherent-detection performance. In this work, we instead use asymmetric constellations as an intrinsic phase reference. We will elaborate on the design of asymmetric constellations in the following section.



\section{Asymmetric Constellation}
\label{sec:asym_constellation}

The blind sensing stage in Sec.~\ref{sec:proposed_method} and the ZF-based spatial separation in Sec.~\ref{sec:data_recovery} facilitate the recovery of source-dependent channel parameters
\(\{(\hat{\tau}_p,\hat{\nu}_p,\hat{\theta}_p,\hat{\alpha}_p)\}_{p=0}^{P-1}\)
and produce per-stream symbol estimates \(\hat{x}_p[k,n]\) on each TF bin \((k,n)\).
However, the fully blind pipeline still leaves ambiguities that are immaterial for sensing, but critical for coherent demodulation:
(i) a constant complex ambiguity (rotation/scale) per recovered stream and (ii) a possible permutation of stream indices.
This section shows that these ambiguities can be resolved blindly by adopting an asymmetric constellation and performing minimum
constellation fitting after the ZF separation. 

\subsection{Asymmetric Fourth-Order Model}
\label{subsec:asym_model}

\subsubsection{Impact on the Fourth-Order Sensing Model}
Let \(\mathcal{C}\subset\mathbb{C}\) denote the constellation used by each source, and assume i.i.d.\ data symbols
\(X_{p,k}[n]\in\mathcal{C}\) with \(\mathbb{E}[|X|^2]=\sigma_x^2\). Let us define the fourth moment as
\begin{equation}
\mu_4 \triangleq \mathbb{E}[X^4] = \sum_{c\in\mathcal{C}} \Pr(X=c)\,c^4 \neq 0.
\end{equation}
In contrast to constellations having constant \(X^4\) (e.g., unit-power QPSK), a generic asymmetric constellation does not satisfy a per-symbol identity \(X^4=\mathrm{const}\).
Instead, we decompose
\begin{equation}
X_{q,k}[n]^4 = \mu_4 + \widetilde{X}^{\,4}_{q,k}[n], \qquad \mathbb{E}\!\left[\widetilde{X}^{\,4}_{q,k}[n]\right]=0,
\end{equation}
where \(\widetilde{X}^{\,4}_{q,k}[n]\) is a zero-mean residual modulation term. Consequently, the coherent component of the fourth-order virtual tensor remains proportional to \(\mu_4\) and preserves the separable
3D-exponential peak structure used by the HOS periodogram; the only constellation-dependent quantity entering the gain recovery is \(\mu_4\) in \eqref{eq:complex_gain_est}.

\subsubsection{Post-ZF Constant-Ambiguity Model}
Using \(\{(\hat{\tau}_p,\hat{\nu}_p,\hat{\theta}_p,\hat{\alpha}_p)\}\), we construct \(\hat{\mathbf A}[k,n]\) as in \eqref{eq:Mixing_a}--\eqref{eq:Mixing_A} and apply the ZF equalizer in \eqref{eq:ZF},
yielding \(\hat{\mathbf x}[k,n]\in\mathbb{C}^{P}\) for each TF bin.
Let \(\hat{x}_p[k,n]\) denote the \(p\)-th entry of \(\hat{\mathbf x}[k,n]\).

Because the fourth-order gain recovery admits a quadrant phase ambiguity and ZF separation preserves a column scaling ambiguity,
the ZF outputs obey the ambiguity model given by
\begin{equation}
\hat{x}_p[k,n] = \beta_p\, X_{\pi(p),k}[n] + \eta_p[k,n], \qquad p=0,\ldots,P-1,
\label{eq:ZF_new}
\end{equation}
where \(\pi(\cdot)\) is an unknown permutation over \(\{0,\ldots,P-1\}\), \(\beta_p\in\mathbb{C}\) is an unknown constant complex factor (rotation/scale),
and \(\eta_p[k,n]\) is the effective post-ZF noise/interference residue.

\subsubsection{Identifiability Enabled by Asymmetry}
If \(\mathcal{C}\) is invariant under a nontrivial rotation (e.g., QPSK where \(\mathcal{C}=e^{j\pi/2}\mathcal{C}\)),
then the phase of \(\beta_p\) in \eqref{eq:ZF_new} is not identifiable from data: the pairs \((\beta_p,X)\) and \((\beta_p e^{-j\varphi},e^{j\varphi}X)\)
produce identical observations whenever \(e^{j\varphi}\mathcal{C}=\mathcal{C}\).
We therefore employ an asymmetric constellation satisfying
\begin{equation}
e^{j\varphi}\mathcal{C}=\mathcal{C}\ \Rightarrow\ \varphi=0,
\end{equation}
which provides an intrinsic phase reference and makes \(\arg(\beta_p)\) uniquely determined (up to noise) by constellation consistency.

\subsubsection{Minimum Constellation Fitting}
Define the nearest-neighbor slicer for \(\mathcal{C}\) as
\begin{equation}
\mathcal{Q}_{\mathcal{C}}(z)\triangleq \arg\min_{c\in\mathcal{C}} |z-c|^2.
\end{equation}
For a candidate permutation \(\pi\), we estimate \(\beta_p\) for each stream by minimizing the total distance between the ZF samples and a scaled/rotated
constellation.
Consider stream \(p\) and the sample set \(\mathcal{S}_p\triangleq\{\hat{x}_p[k,n]\}_{k=0,n=0}^{K-1,\,N_{\mathrm{sym}}-1}\).
We then solve the following joint least-squares (LS) fitting problem
\begin{equation}
\begin{aligned}
    \big(\hat{\beta}_p,\{\hat{X}_{\pi(p),k}[n]\}\big)\in
\arg\min_{\beta\in\mathbb{C},\,X_{k}[n]\in\mathcal{C}}
\sum_{k=0}^{K-1}\sum_{n=0}^{N_{\mathrm{sym}}-1} \\ \left|\hat{x}_p[k,n]-\beta X_{k}[n]\right|^2.
\end{aligned}
\end{equation}
For a fixed \(\beta\), the minimizer is obtained by slicing \(\hat{X}_{k}[n]=\mathcal{Q}_{\mathcal{C}}(\hat{x}_p[k,n]/\beta)\).
Substituting back yields a scalar-only objective
\begin{equation}
\hat{\beta}_p(\pi)\in \arg\min_{\beta\in\mathbb{C}}
\sum_{k=0}^{K-1}\sum_{n=0}^{N_{\mathrm{sym}}-1}
\left|\hat{x}_p[k,n]-\beta\,\mathcal{Q}_{\mathcal{C}}\!\big(\hat{x}_p[k,n]/\beta\big)\right|^2.
\label{eq:constellation_matching_obj}
\end{equation}
We solve (\ref{eq:constellation_matching_obj}) via a coarse phase search for initialization, followed by a few decision-directed LS refinements. Starting from
\(\beta_p^{(0)}\), we iterate for \(t=0,1,\ldots,T-1\):
\begin{equation}
\hat{X}^{(t)}_{\pi(p),k}[n] = \mathcal{Q}_{\mathcal{C}}\!\left(\hat{x}_p[k,n]/\beta^{(t)}_p\right),
\end{equation}
\begin{equation}
\beta^{(t+1)}_p =
\frac{\sum_{k,n} \hat{x}_p[k,n]\left(\hat{X}^{(t)}_{\pi(p),k}[n]\right)^{\!*}}
{\sum_{k,n}\left|\hat{X}^{(t)}_{\pi(p),k}[n]\right|^2+\varepsilon},
\end{equation}
where \(\varepsilon>0\) is a small regularizer.
The fitting residual for stream \(p\) is then given by
\begin{equation}
J_p(\pi)\triangleq \sum_{k,n}\left|\hat{x}_p[k,n]-\hat{\beta}_p(\pi)\,\hat{X}_{\pi(p),k}[n]\right|^2,
\end{equation}
where \(\hat{X}_{\pi(p),k}[n]\) denotes the final sliced symbols.

\subsubsection{Permutation Selection and Blind Demodulation}
We select the specific permutation that minimizes the total fitting residual:
\begin{equation}
\hat{\pi}\in \arg\min_{\pi\in \mathcal{S}_P}\ \sum_{p=0}^{P-1} J_p(\pi),
\end{equation}
where \(\mathcal{S}_P\) is the set of all permutations of \(P\) elements.
Finally, blind coherent demodulation is obtained by per-stream alignment and slicing:
\begin{equation}
\widetilde{X}_{\hat{\pi}(p),k}[n] = \mathcal{Q}_{\mathcal{C}}\!\left(\hat{x}_p[k,n]/\hat{\beta}_p(\hat{\pi})\right),
\qquad \forall\,k,n,\ \forall\,p.
\end{equation}

\subsection{Asymmetric Constellation Design}
\label{subsec:asym_design}


This subsection provides a practical design rubric for an asymmetric constellation \(\mathcal{C}\) that (i) preserves communication performance, (ii) maintains the fourth-order sensing model through \(\mu_4\neq 0\), and (iii) resolves the residual rotation/scale/permutation ambiguities by breaking rotational symmetry. Although the criteria apply to a general asymmetric constellation, we adopt amplitude-phase shift keying (APSK)~\cite{1092165} as the design family in this work, since its independently tunable ring radii and phase offsets can break rotational symmetry while maintaining a relatively high minimum Euclidean distance. APSK also supports lookup-table-based implementation, making it a practical baseband modification for OFDM systems.

\subsubsection{Communication Criterion}
Let \(\mathcal{C}=\{c_1,\ldots,c_{|\mathcal{C}|}\}\subset\mathbb{C}\). We adopt the minimum squared Euclidean distance (MSED)
\begin{equation}
d_{\min}^2(\mathcal{C}) \triangleq \min_{i\neq j} |c_i-c_j|^2,
\end{equation}
as the primary metric governing the nearest-neighbor error performance. Under the average-power normalization of
\begin{equation}
\mathbb{E}[|X|^2]
=\sum_{c\in\mathcal{C}}\Pr(X=c)|c|^2
=\sigma_x^2,
\end{equation}
a higher \(d_{\min}^2(\mathcal{C})\) generally yields a lower symbol error probability. In the simulations, the symbols are drawn uniformly from \(\mathcal{C}\), so \(\Pr(X=c)=1/|\mathcal{C}|\), and each candidate APSK constellation is normalized to unit average energy before transmission.

\subsubsection{Sensing Constraint}
To preserve the coherent peak structure in the fourth-order domain and the gain relationship, the constellation must satisfy
\begin{equation}
\mu_4 \triangleq \mathbb{E}[X^4]
=\sum_{c\in\mathcal{C}} \Pr(X=c)\,c^4 \neq 0.
\end{equation}
Since \(\mu_4\) depends only on \(\mathcal{C}\) and its symbol probabilities, it can be computed offline and substituted into \eqref{eq:complex_gain_est} without changing Algorithm~\ref{alg:blind_hos}.

In addition, because the sensing front-end is based on the fourth-order statistic \(E[Y^4]\), the expansion of \(Y_{k,n,m}^4\) contains mixed even-power inter-source terms whose expectation depends on \(\mathbb{E}[X^2]\). If \(\mathbb{E}[X^2]\neq 0\), these terms may produce additional coherent components and hence spurious peaks in the HOS periodogram. Therefore, besides requiring \(\mu_4\neq 0\), the asymmetric constellation should keep the lower-order moments \(|\mathbb{E}[X]|\) and \(|\mathbb{E}[X^2]|\) as small as possible. This is enforced at the design stage through a soft-penalty criterion rather than an exact equality constraint, which provides greater flexibility for maintaining both HOS robustness and communication distance.

\subsubsection{Blind Identifiability}
Blind demodulation requires that \(\mathcal{C}\) have no nontrivial rotational invariance. Equivalently, we have
\begin{equation}
e^{j\varphi}\mathcal{C}=\mathcal{C}\ \Rightarrow\ \varphi=0.
\end{equation}
Beyond this binary condition, the degree of symmetry breaking strongly impacts robustness at finite SNR. We quantify it through the rotation-separation distance:
\begin{equation}
d_{\mathrm{rot}}(\varphi;\mathcal{C})
\triangleq
\frac{1}{|\mathcal{C}|}\sum_{c\in\mathcal{C}}
\min_{c'\in\mathcal{C}}
\left|c-e^{j\varphi}c'\right|,
\label{eq:drot_new}
\end{equation}
which measures how far the rotated constellation \(e^{j\varphi}\mathcal{C}\) lies from the original set. In particular, the quadrant ambiguity imposed by the fourth-root phase uncertainty motivates requiring
\begin{equation}
d_{\mathrm{rot}}\!\left(\frac{\pi}{2};\mathcal{C}\right) > 0.
\end{equation}

\subsubsection{Design Objective}
Combining the above requirements, we adopt the following design principle: a good asymmetric constellation should have high \(d_{\min}(\mathcal{C})\), high \(|\mu_4|\), and high rotational separation, while keeping \(|\mathbb{E}[X]|\) and \(|\mathbb{E}[X^2]|\) small. A scalar objective is therefore formulated as
\begin{equation}
\begin{aligned}
\max_{\mathcal{C}}
\quad
w_d\, d_{\min}(\mathcal{C})
+ w_4\, |\mathbb{E}[X^4]|
+ w_r\, d_{\mathrm{rot}}\!\left(\frac{\pi}{2};\mathcal{C}\right)
- \\ w_1\, |\mathbb{E}[X]|^2
- w_2\, |\mathbb{E}[X^2]|^2,
\label{eq:const_design_obj}
\end{aligned}
\end{equation}
where \(w_d,w_4,w_r,w_1,w_2>0\) are design weights. 

\subsubsection{Ambiguity Resistance}
Under the constant-ambiguity model in \eqref{eq:ZF_new}, incorrect rotations incur a systematic mismatch governed by the separation between \(\mathcal{C}\) and its rotated copies, while noise imposes fluctuations on the order of \(N_0\). Since the fitting metric aggregates \(L\approx KN_{\mathrm{sym}}\) samples, a sufficient reliability condition is
\begin{equation}
|\beta_p|^2\, d_{\mathrm{rot}}^2\!\left(\frac{\pi}{2};\mathcal{C}\right)
\gg \frac{N_0}{L},
\qquad
L \approx K N_{\mathrm{sym}}.
\end{equation}
This highlights an important ISAC advantage: because OFDM frames provide substantial averaging over the TF plane, only mild geometric asymmetry is typically needed to suppress ambiguity occurrences while retaining near-optimal communication performance.

In summary, the asymmetric constellation should satisfy the following condition:
maximize \(d_{\min}^2(\mathcal{C})\) under \(\mathbb{E}[|X|^2]=\sigma_x^2\), \(\mu_4\neq 0\), \(d_{\mathrm{rot}}(\pi/2;\mathcal{C})>0\),
and preferably
\begin{equation}
|\beta_p|^2\, d_{\mathrm{rot}}^2\!\left(\frac{\pi}{2};\mathcal{C}\right)\gg \frac{N_0}{K N_{\mathrm{sym}}}.
\end{equation}
This rubric preserves MSED-driven communication performance, while suppressing ambiguity-induced error floors, enabling robust blind demodulation.

\section{Performance Analysis: The Cost of Blindness}
\label{sec:CRLB}

To benchmark the proposed blind estimator, we compare its MSE against the theoretical fundamental limits. We consider two bounds: the data-aided CRLB, which assumes perfect knowledge of the transmitted symbols, and the stochastic (blind) CRLB, which treats the data as unknown random nuisance contributions.

    

    

\subsection{Analytical Lower Bounds}
Let the unknown parameter vector for a single source be $\bm{\xi} = [\tau, \nu, \theta]^T$. As derived in Appendix \ref{app:crlb_derivation}, the data-aided CRLB for each parameter is given by
\begin{align}
    \text{CRLB}_{DA}(\tau) &= \frac{1}{2\cdot \text{SNR} \cdot G_{tot} \cdot \beta_f^2}, \label{eq:crb_tau} \\
    \text{CRLB}_{DA}(\nu) &= \frac{1}{2\cdot \text{SNR} \cdot G_{tot} \cdot \beta_t^2} \cdot \left( \frac{c}{f_c} \right)^2, \label{eq:crb_vel} \\
    \text{CRLB}_{DA}(\theta) &= \frac{1}{2\cdot \text{SNR} \cdot G_{tot} \cdot \beta_s^2} \cdot \left( \frac{180}{\pi} \right)^2, \label{eq:crb_ang} 
\end{align}
where $\text{SNR} = \sigma_x^2 / \sigma_n^2$ is the linear signal-to-noise ratio, and $G_{tot} = K N_{sym} M$ is the total processing gain of the 3D data cube. The terms $\beta_f^2, \beta_t^2, \beta_s^2$ represent the effective mean-square bandwidths in the frequency, time, and spatial domains, respectively:
\begin{equation}
    \beta_f^2 = (2\pi \Delta f)^2 \frac{K^2-1}{12}, 
\end{equation}
\begin{equation}
    \beta_t^2 = (2\pi T_{sym})^2 \frac{N_{sym}^2-1}{12},
\end{equation}
\begin{equation}
    \beta_s^2 = \left(2\pi \frac{d}{\lambda}\right)^2 \frac{M^2-1}{12}.
\end{equation}

\subsection{Asymptotic Convergence of Blind and Data-Aided Bounds}
A key theoretical insight derived in Appendices \ref{app:stochastic_crb} is that for high-SNR frequency estimation, the stochastic CRLB converges to the data-aided CRLB:
\begin{equation}
    \lim_{\text{SNR} \to \infty} \frac{\text{CRLB}_{Blind}(\bm{\xi})}{\text{CRLB}_{DA}(\bm{\xi})} = 1.
\end{equation}
This implies that, theoretically, the penalty for not knowing the data vanishes, if the optimal blind estimator is used.

\begin{figure}[!t]
    \centering
    \subfloat[Range-velocity heatmap.]
    {
        \label{fig:range_velocity_heatmap}
        \includegraphics[width=1.62in]{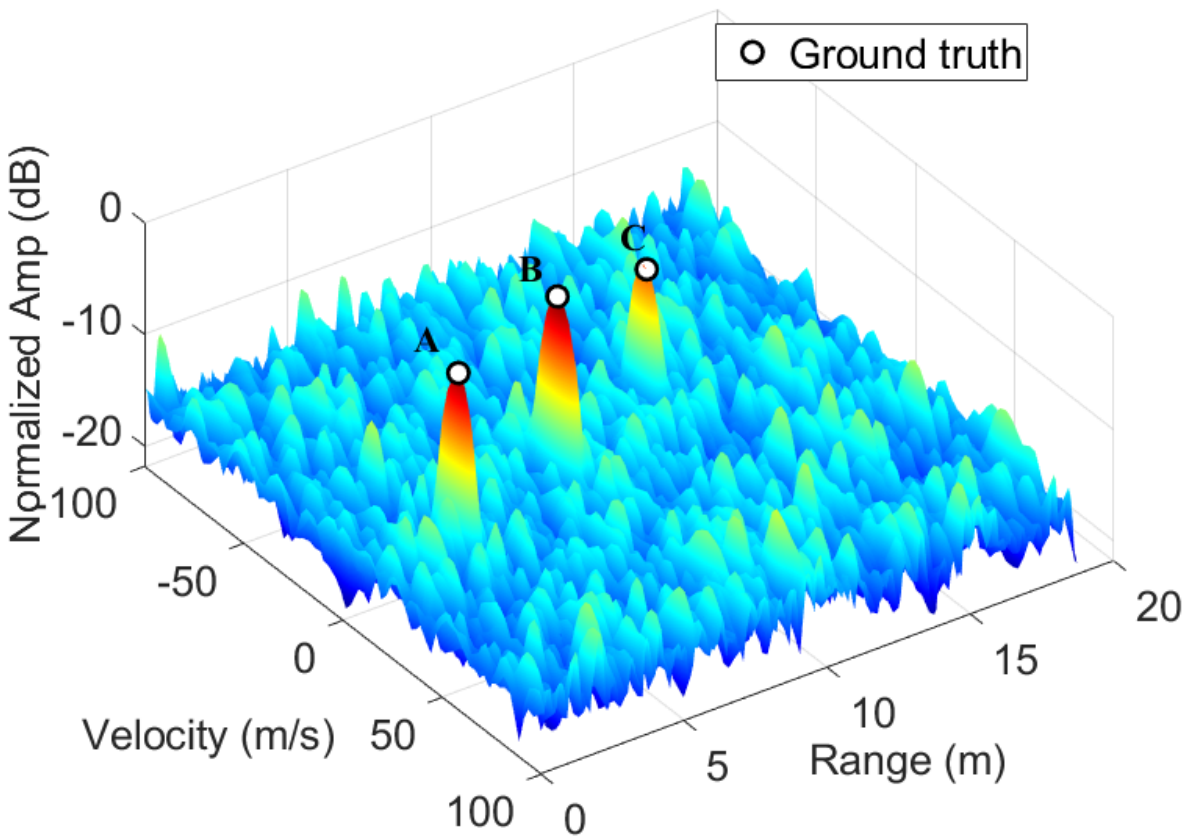}
    }
    \hspace{0.0001\linewidth}
    \subfloat[Range-angle heatmap.]
    {
       \label{fig:range_angle_heatmap}
        \includegraphics[width=1.62in]{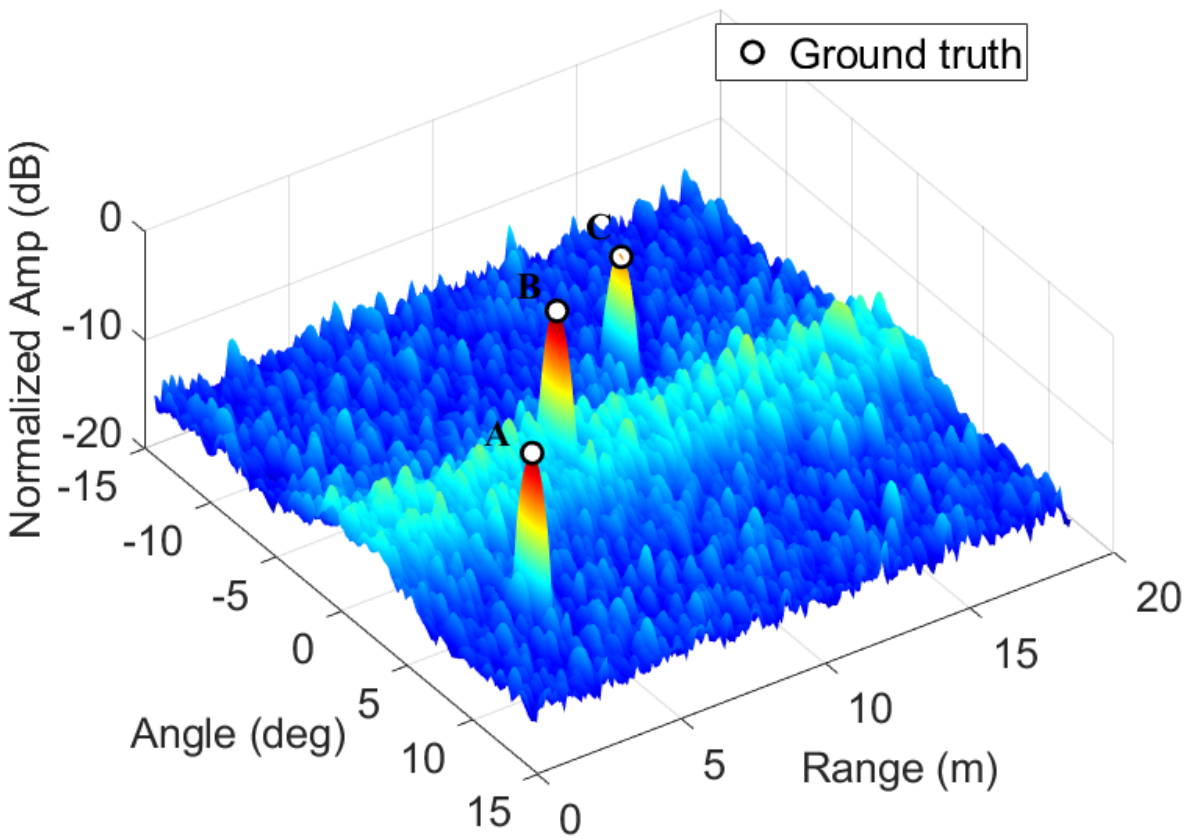}
    }
    
\caption{Simulation results of estimations}
\label{fig:sim_heatmap}
\end{figure}

\begin{figure*}[!t]
    \centering
    \subfloat[RMSE of delay estimation versus SNR.]{
        \includegraphics[width=1.8in]{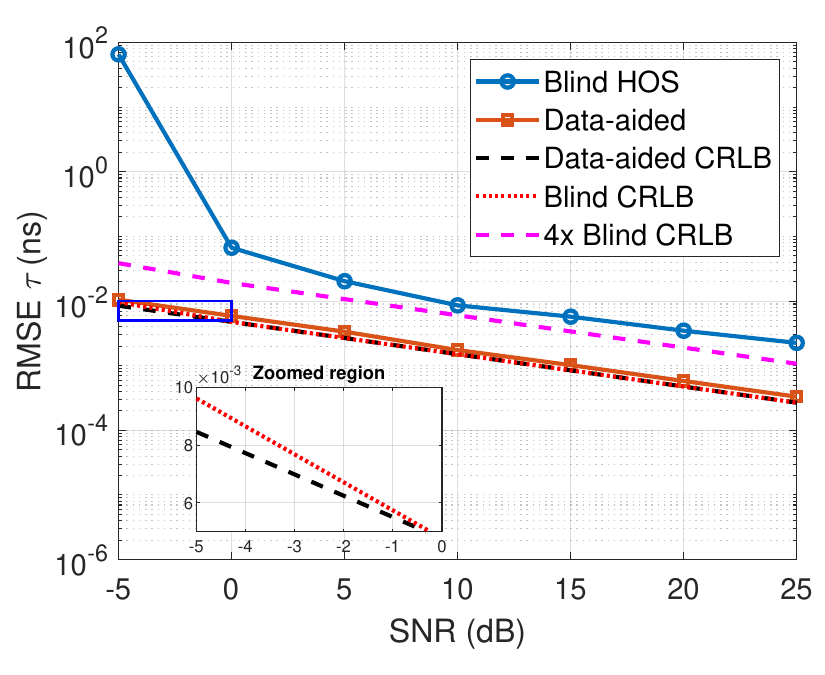}
        \label{fig:RMSE_CRLB_delay}
    }
    \hspace{0.01\linewidth}
    \subfloat[RMSE of velocity estimation versus SNR.]{
        \includegraphics[width=1.8in]{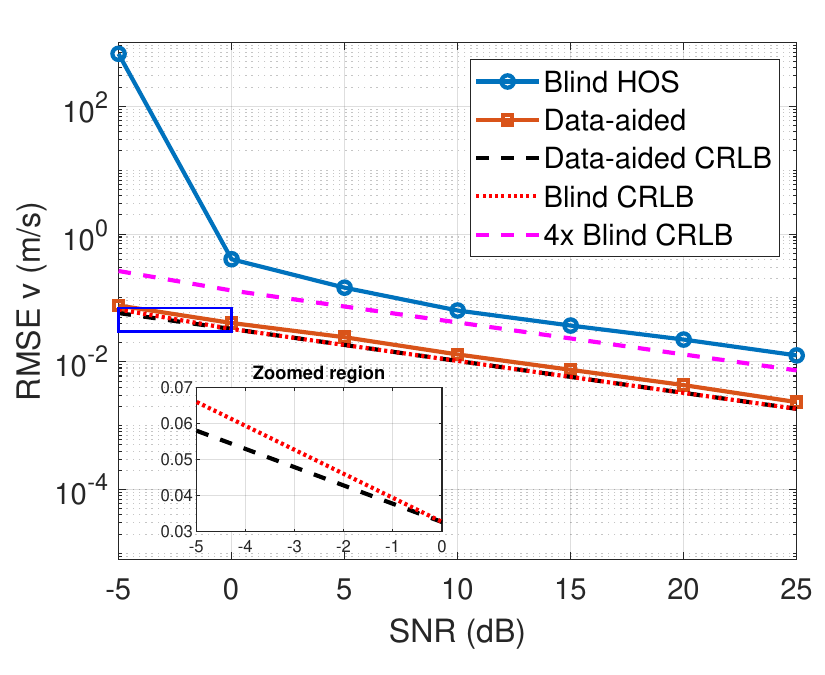}
        \label{fig:RMSE_CRLB_velocity}
    }
    \hspace{0.01\linewidth}
    \subfloat[RMSE of angle estimation versus SNR.]{
        \includegraphics[width=1.8in]{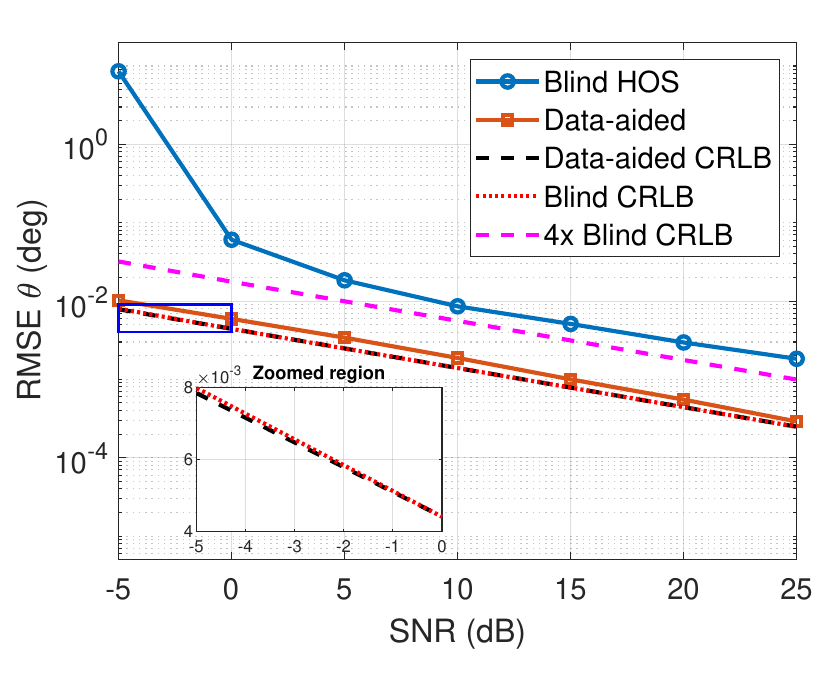}
        \label{fig:RMSE_CRLB_angle}
    }

    \subfloat[RMSE of delay estimation versus number of pilot subcarriers.]{
        \includegraphics[width=1.8in]{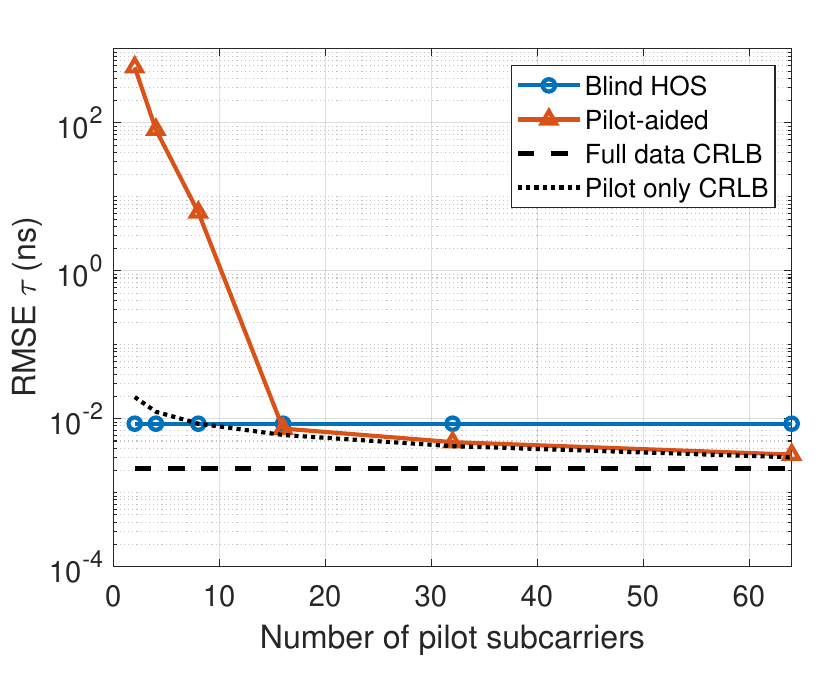}
        \label{fig:RMSE_pilot_delay}
    }
    \hspace{0.01\linewidth}
    \subfloat[RMSE of velocity estimation versus number of pilot subcarriers.]{
        \includegraphics[width=1.8in]{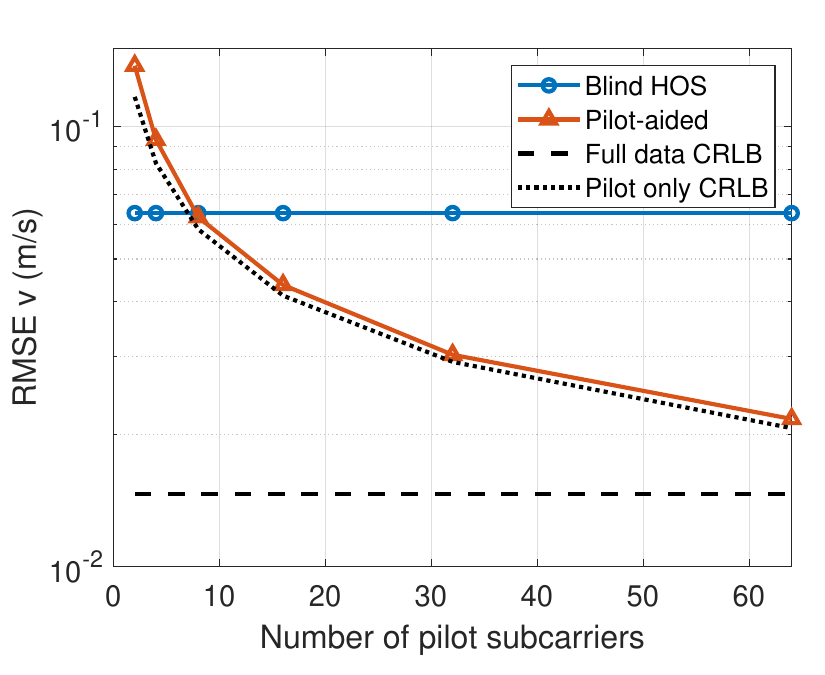}
        \label{fig:RMSE_pilot_velocity}
    }
    \hspace{0.01\linewidth}
    \subfloat[RMSE of angle estimation versus number of pilot subcarriers.]{
        \includegraphics[width=1.8in]{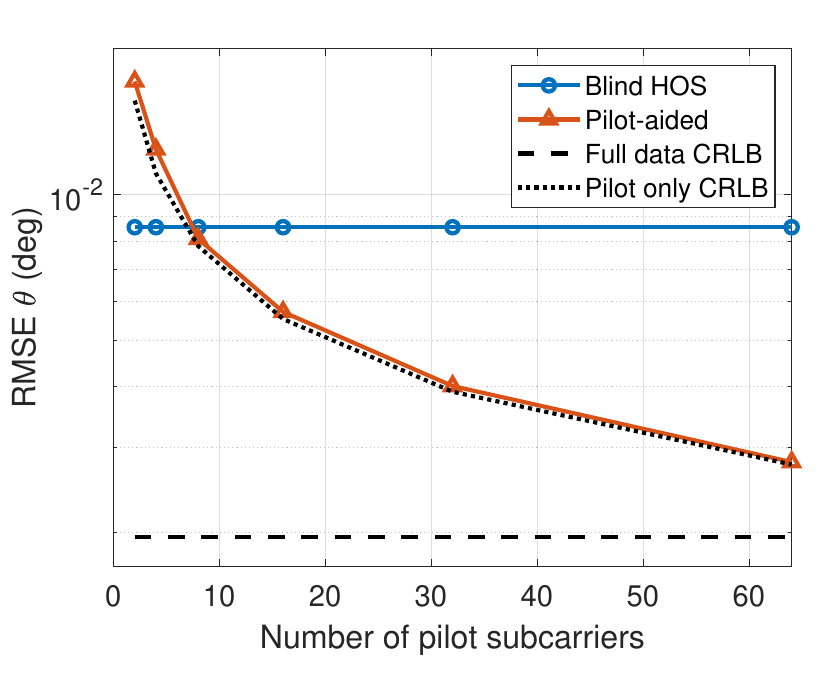}
        \label{fig:RMSE_pilot_angle}
    }

    \caption{RMSE performance of the proposed blind HOS estimator and benchmark methods. The top row shows delay, velocity, and angle estimation errors versus SNR. The bottom row shows the corresponding RMSE versus the number of pilot subcarriers.}
    \label{fig:RMSE_all}
\end{figure*}

\begin{figure}[!t]
    \centering
    \subfloat[Sensing prioritized constellation.]
    {
        \label{fig:sync_timing}
        \includegraphics[width=1.0in]{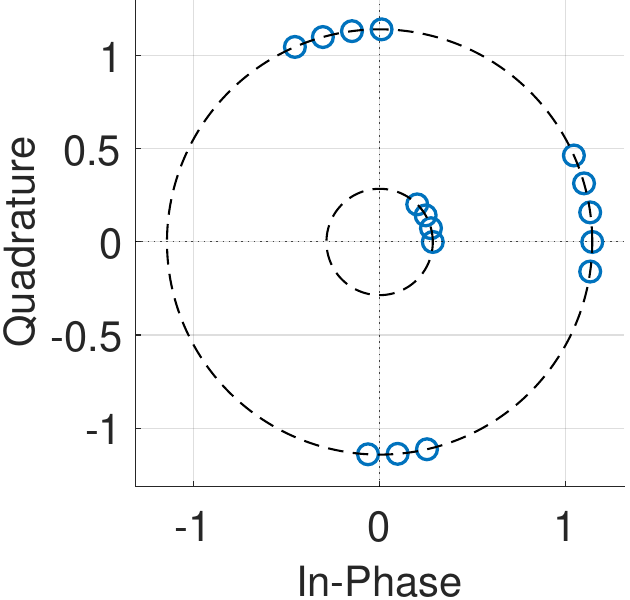}
    }
    \hspace{0.0001\linewidth}
    \subfloat[Communication prioritized constellation.]
    {
       \label{fig:sync_cfo}
        \includegraphics[width=1.in]{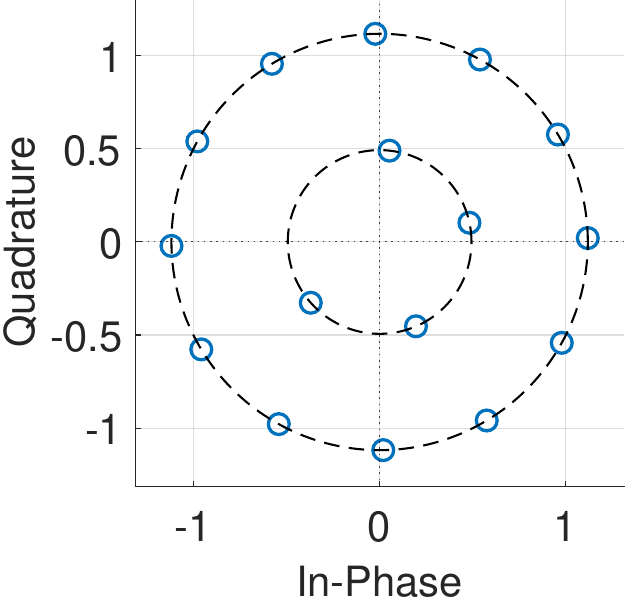}
    }
    \hspace{0.0001\linewidth}
    \subfloat[Balanced constellation.]
    {
        \label{fig:sync_phase}
        \includegraphics[width=1.0in]{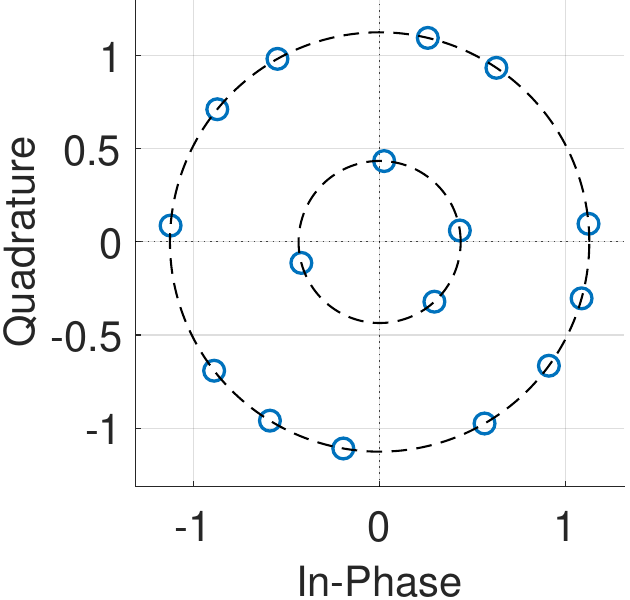}
    }
    
\caption{Asymmetric APSK constellation designs}
\label{fig:constellation_APSK}
\end{figure}

\subsection{The Cost of Blindness}\label{sec:cost}

Although the stochastic CRLB approaches the data-aided CRLB at high SNR, the proposed blind estimator still incurs a practical performance penalty relative to an ideal data-aided receiver. This penalty does not arise because the sensing parameters become fundamentally unidentifiable, but because the blind front-end harnessed relies on a fourth-order transformation of the received data, which amplifies noise and residual cross terms.

{To see this effect, we write the received sample as \(Y=S+W\), where $S$ denotes the noiseless signal component and $W$ denotes additive noise. Under a low-noise approximation, the fourth-power operation yields
\begin{equation}
(S+W)^4 \approx S^4 + 4S^3W.
\end{equation}
Hence, the leading perturbation term after fourth-order processing is scaled by $4S^3$, so its variance is amplified by approximately $|4S^3|^2 = 16|S|^6$.
Under unit-amplitude normalization, i.e., $|S|\approx 1$, this corresponds to an approximately 16-fold increase in the variance of the leading perturbation term.}

Therefore, the practical cost of blindness in the proposed HOS framework should be interpreted as an implementation loss caused by fourth-order processing, rather than as a fundamental loss of identifiability. In particular, although the blind and data-aided CRLBs become asymptotically identical at high SNR, the proposed estimator generally requires additional averaging or coherent integration gain, such as a higher number of OFDM symbols $N_{\mathrm{sym}}$, in order to overcome the increased noise sensitivity introduced by the fourth-order transformation.

\begin{table*}[!t]
\centering
\caption{Representative asymmetric 16-APSK designs.}
\label{tab:constellation_design}
\begin{tabular}{c|ccccc|cccccc}
\hline
 & $w_d$ & $w_4$ & $w_r$ & $w_1$ & $w_2$ & $r_2$ & $d_{\min}(\mathcal C)$ & $|\mathbb E[X^4]|$ & $d_{\rm rot}(\pi/2;\mathcal C)$ & $\frac{E_{\rm peak}}{\mathbb E[|X|^2]}$ \\
\hline
Balanced & 190 & 60 & 25 & 200 & 700 &  2.588 & 0.402 & 0.277 & 0.154 & 1.270 \\
Communication-prioritized & 600 & 5 & 80 & 200 & 120 &  2.272 & 0.579 & 0.004 & 0.064 & 1.252 \\
Sensing-prioritized & 80 & 180 & 10 & 150 & 1200 &  4.000 & 0.0746 & 0.999 & 0.447 & 1.306 \\
\hline
\end{tabular}
\end{table*}

\begin{figure*}[!t]
    \centering
    \subfloat[Communication performance.]
    {
        \label{fig:performance_constellation_designs1}
        \includegraphics[width=1.6in]{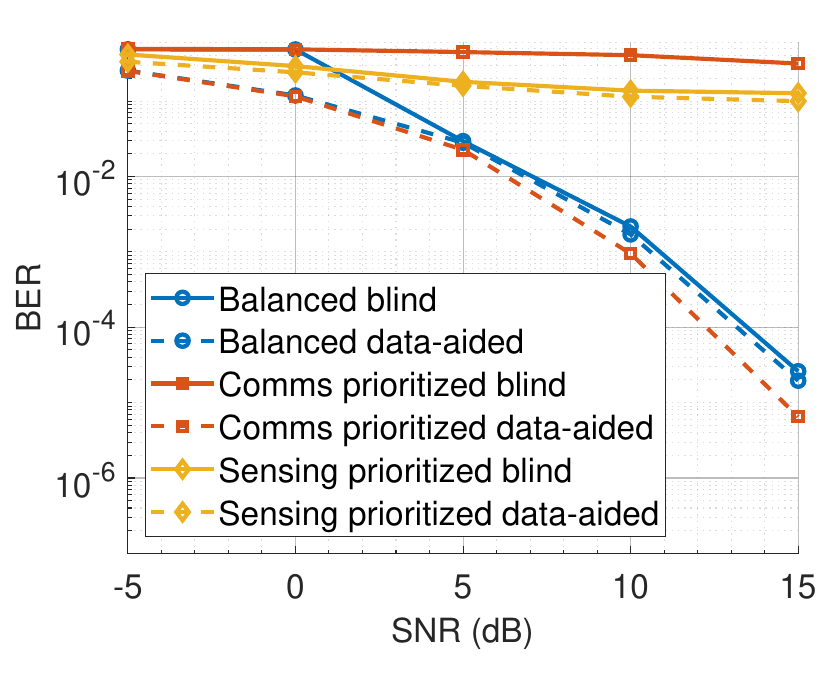}
    }
    \hspace{0.0001\linewidth}
    \subfloat[Delay estimation performance.]
    {
       \label{fig:performance_constellation_designs2}
        \includegraphics[width=1.6in]{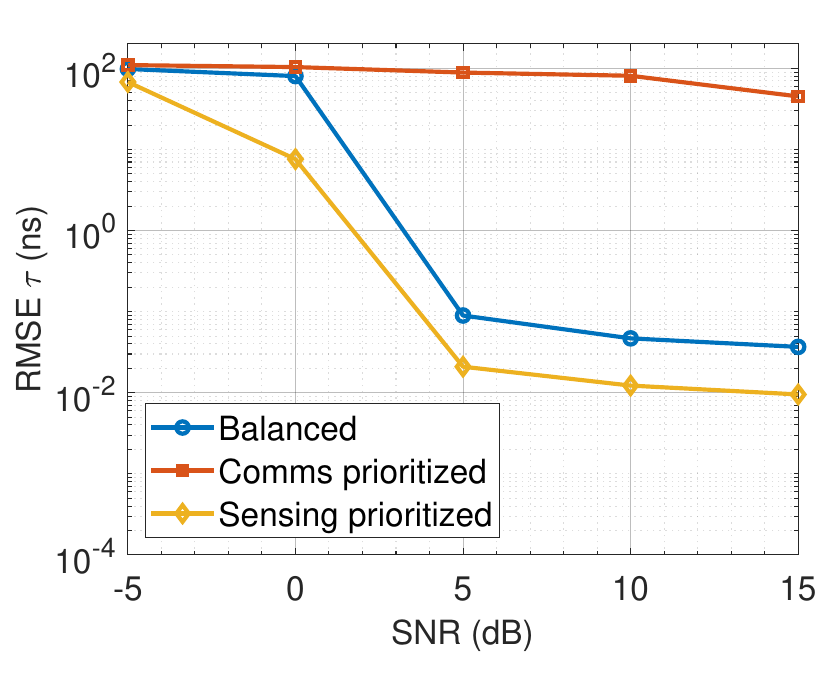}
    }
    \hspace{0.0001\linewidth}
    \subfloat[Velocity estimation performance.]
    {
        \label{fig:performance_constellation_designs3}
        \includegraphics[width=1.6in]{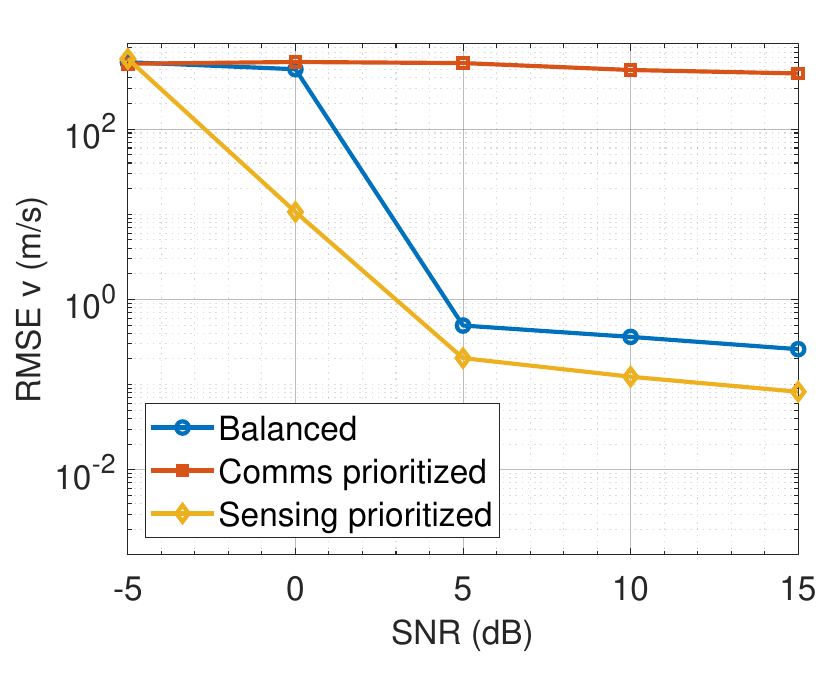}
    }
    \hspace{0.0001\linewidth}
    \subfloat[Angle estimation performance.]
    {
        \label{fig:performance_constellation_designs4}
        \includegraphics[width=1.6in]{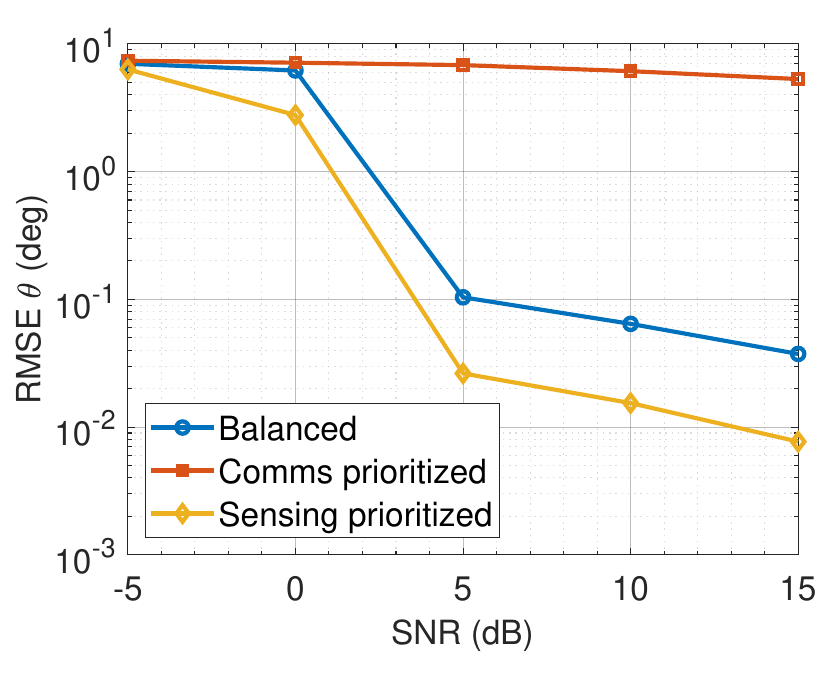}
    }
    
\caption{Communication and sensing performance with different constellation designs.}
\label{fig:performance_constellation_designs}
\end{figure*}

\section{Numerical and Experimental Results}
\label{sec:results}

In this section, we validate the proposed HOS-based blind sensing framework through extensive numerical simulations and hardware experiments. 

\subsection{Simulation Setup}

In the simulations, we consider an OFDM-based ISAC receiver equipped with a ULA of \(M=16\) antennas operating at a carrier frequency of \(f_c=28~\mathrm{GHz}\). The OFDM waveform employs \(N=128\) subcarriers with useful symbol duration of \(T_u=1~\mu\mathrm{s}\), which yields a subcarrier spacing of \(\Delta f = 1/T_u\), and we set the OFDM symbol interval to \(T_{\mathrm{sym}}=T_u\). The total signal bandwidth is 128~MHz. The HOS is calculated using \(N_{\mathrm{sym}}=200\) OFDM symbols. Unless otherwise specified, the following results are presented with the above setting.

\subsection{Blind Sensing Performance}
We first visualize the proposed blind estimator using 3D heatmaps over both the range--velocity and range--angle domains. The received signal is modeled as a superposition of \(P=3\) sources with \((\tau_p, v_p, \theta_p)\) given by \((20~\mathrm{ns},\,15~\mathrm{m/s},\,8^\circ)\), \((80~\mathrm{ns},\,-20~\mathrm{m/s},\,-5^\circ)\), and \((50~\mathrm{ns},\,0~\mathrm{m/s},\,0^\circ)\), respectively, and complex amplitudes of \(|\alpha_p|\in\{1.0,\,0.9,\,1.0\}\). The received SNR is set to \(5~\mathrm{dB}\). Unless otherwise specified, QPSK is used in this subsection. Figures~\ref{fig:range_velocity_heatmap} and \ref{fig:range_angle_heatmap} show the resultant range–velocity and range–angle maps, respectively. As observed, the dominant peaks in the heatmaps closely match the ground-truth source locations.

We next evaluate the estimation accuracy of the sensing parameters, namely delay, velocity, and angle. The root-mean-square error (RMSE) is adopted as the performance metric, and the proposed HOS-based blind estimator is compared to a data-aided baseline as well as the corresponding CRLBs derived. The results are reported in Fig.~\ref{fig:RMSE_all} (a)--(c). As shown, the data-aided estimator closely attains the CRLB for all three parameters. By contrast, the blind HOS estimator exhibits an approximately four-fold performance loss relative to the CRLB, which is consistent with the analysis in Section~\ref{sec:cost}.

To further evaluate the proposed blind sensing method, we compare it to pilot-aided channel estimation under a comb-pilot configuration, where the number of pilot subcarriers varies from $2$ to $64$ out of $128$ subcarriers. The results at an SNR of $10$ dB are shown in Fig.~\ref{fig:RMSE_all}(d)--(f). When the pilot density is low, the blind HOS estimator achieves better accuracy. In particular, it consistently outperforms the pilot-aided scheme in delay estimation, when no more than $16$ pilot subcarriers are used, corresponding to a pilot overhead of $12.5\%$. For velocity and angle estimation, it also performs better when the number of pilots is extremely limited, for example with only $4$ pilot subcarriers. This is because pilot-aided estimation relies only on pilot observations, so sparse pilots reduce the sensing information and degrade accuracy. By contrast, the blind HOS method exploits the entire received data block and only incurs the four-fold processing loss discussed in Section~\ref{sec:cost}. In a comb-pilot setting, delay estimation is particularly sensitive to pilot spacing across frequency, so the pilot density has the strongest impact on delay accuracy. Similar behavior is expected for other pilot patterns, such as block-type pilots, where limited pilot symbols would more strongly affect velocity estimation. Overall, pilot-aided estimation is only preferable when the pilot density is high enough for ensuring that its performance loss becomes smaller than the $4\times$ penalty of the blind HOS method relative to the CRLB. Otherwise, the blind approach is more attractive, because it avoids pilot overhead and preserves the full communication payload, while still exploiting the entire received signal for sensing.

\subsection{Asymmetric Constellation Design}

To illustrate the design principle formulated in \eqref{eq:const_design_obj}, we optimize a twin-ring 16-APSK constellation by tuning the ring phases and the outer-ring radius. In particular, the design follows the weighted objective in \eqref{eq:const_design_obj}, which balances communication reliability through the minimum Euclidean distance \(d_{\min}(\mathcal C)\), sensing coherence through the fourth moment \(\mathbb E[X^4]\), and ambiguity resistance through the rotational-separation metric \(d_{\rm rot}(\pi/2;\mathcal C)\), while penalizing nonzero lower-order moments. In the numerical study, we consider three representative weight settings in \eqref{eq:const_design_obj}: a balanced design, a communication-prioritized design, and a sensing-prioritized design. All designs use a two-ring APSK geometry with \(M_1=4\) inner-ring points and \(M_2=12\) outer-ring points. After optimization, each constellation is normalized to unity average energy.


\begin{figure*}[!t]
    \centering
    \subfloat[Time offset.]
    {
        \label{fig:sync_timing}
        \includegraphics[width=2.in]{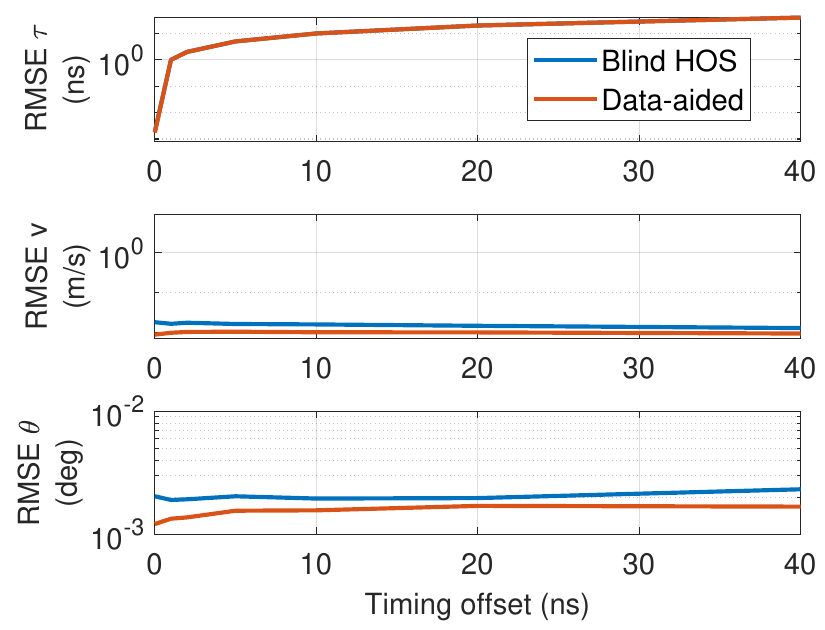}
    }
    \hspace{0.0001\linewidth}
    \subfloat[Carrier frequency offset.]
    {
       \label{fig:sync_cfo}
        \includegraphics[width=2.in]{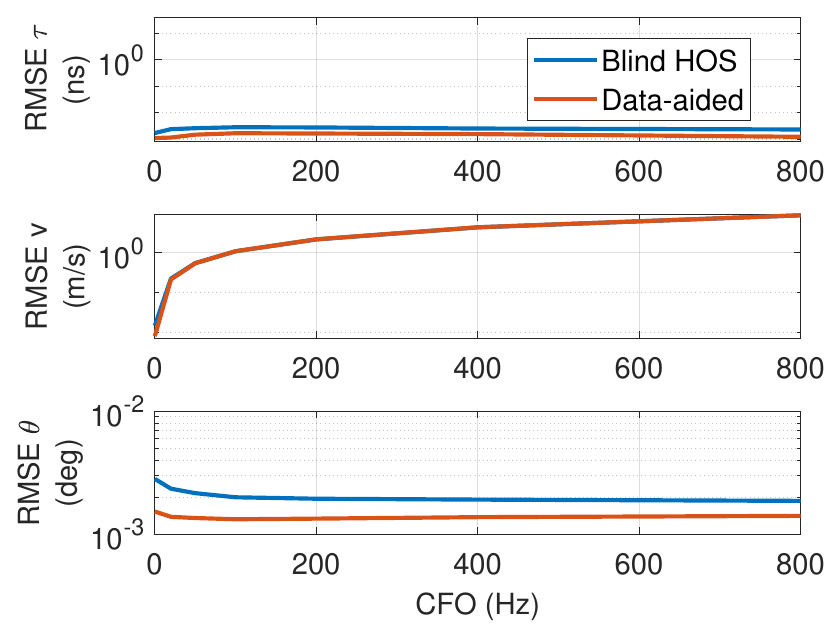}
    }
    \hspace{0.0001\linewidth}
    \subfloat[Phase offset.]
    {
        \label{fig:sync_phase}
        \includegraphics[width=2.in]{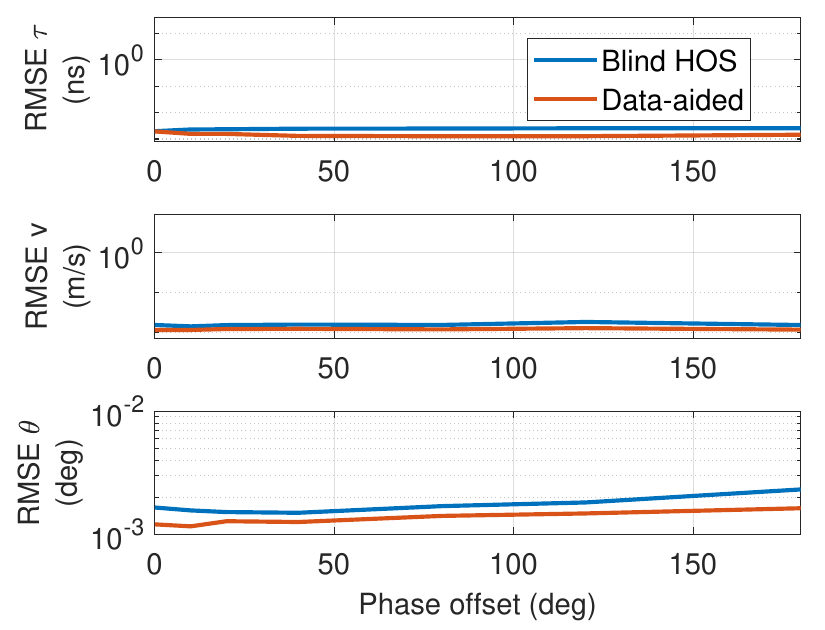}
    }
    
\caption{RMSE performance for delay, velocity, and angle estimation with synchronization impairments.}
\label{fig:sync}
\end{figure*}

\begin{figure}[!t]
    \centering
    \subfloat[Experimental setting.]
    {
        \label{fig:experiment_setting}
        \includegraphics[width=1.7in]{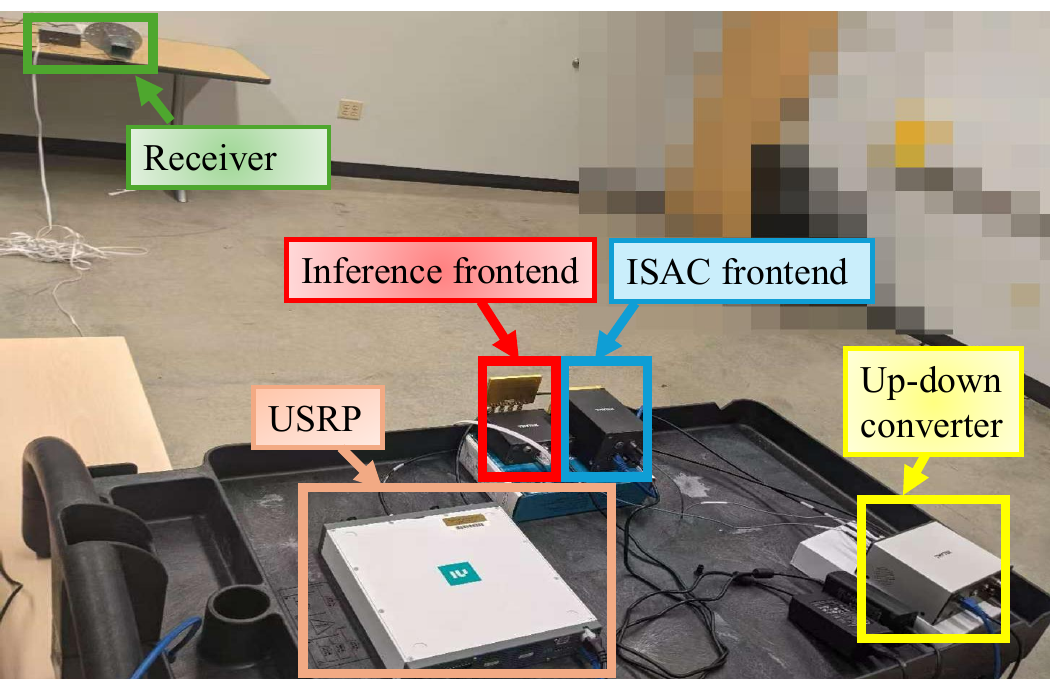}
    }
    \hspace{0.0001\linewidth}
    \subfloat[Experimental results of range-velocity heatmap.]
    {
       \label{fig:exp_range_velocity_heatmap}
        \includegraphics[width=1.5in]{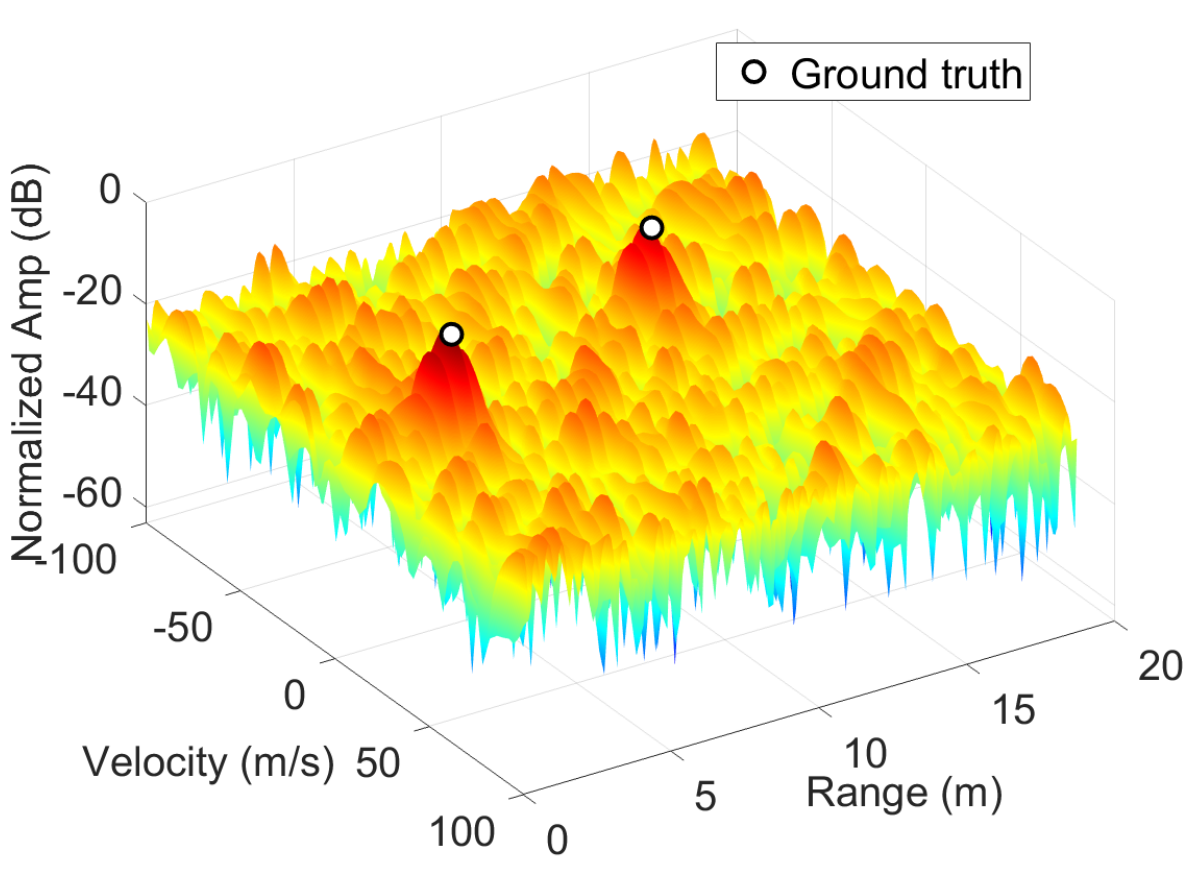}
    }
    
\caption{Simulation results of estimation heatmap}
\label{fig:experiment}
\end{figure}

Table~\ref{tab:constellation_design} summarizes the three representative designs obtained from \eqref{eq:const_design_obj}. Increasing the weight on \(d_{\min}(\mathcal C)\) improves communication robustness by increasing of the minimum distance, but typically reduces \(\mathbb E[X^4]\) and weakens the fourth-order coherence used for blind sensing. Conversely, emphasizing \(\mathbb E[X^4]\) strengthens the HOS sensing structure at the cost of a more compressed constellation geometry and a reduced \(d_{\min}(\mathcal C)\), which degrades demodulation reliability. The balanced design lies between these two extremes. In all cases, \(d_{\rm rot}(\pi/2;\mathcal C)>0\), ensuring distinguishability under the blind rotational ambiguity model. The corresponding constellations are shown in Fig.~\ref{fig:constellation_APSK}.

Fig.~\ref{fig:performance_constellation_designs} compares the communication and sensing performance achieved by the three asymmetric constellation designs. For communication, we also include a data-aided channel-estimation baseline, which serves to reveal the inherent communication capability of each constellation under accurate channel knowledge. As shown in Fig.~\ref{fig:performance_constellation_designs1}, the communication-prioritized design attains the best data-aided BER, consistent with its higher minimum distance \(d_{\min}(\mathcal C)\). By contrast, the sensing-prioritized design exhibits the worst data-aided BER, because its constellation geometry is significantly more compressed, resulting in reduced noise resistance. The balanced design lies between these two extremes. The performance trend under the proposed blind pipeline is more subtle, since blind demodulation depends not only on the intrinsic communication geometry of the constellation, but also on the quality of the channel parameters estimated by the HOS sensing stage. This effect is evident in Figs.~\ref{fig:performance_constellation_designs2}--\ref{fig:performance_constellation_designs4}. Although the communication-prioritized constellation has the most favorable \(d_{\min}(\mathcal C)\), it yields poor delay, velocity, and angle estimation, because its fourth-order coherence and rotational separation is weak. As a result, the reconstructed channel coefficients are inaccurate, and its blind BER becomes substantially worse than that of its data-aided counterpart. By contrast, the sensing-prioritized design delivers the most accurate parameter estimation among the three constellations, which explains why its blind BER closely tracks the corresponding data-aided baseline. Nevertheless, because its minimum distance is inherently small, its absolute BER performance remains limited even when the channel is accurately estimated. The balanced design provides the most favorable overall tradeoff. It preserves sufficient fourth-order structure to enable reliable delay, Doppler, and angle estimation, while still maintaining an adequate minimum distance for symbol detection. Consequently, it achieves competitive sensing accuracy together with strong blind communication performance, and strikes the best end-to-end compromise between the sensing and communication objectives.

\subsection{Impact of Synchronization Impairments}

Finally, we evaluate the impact of synchronization impairments on sensing performance by comparing the proposed blind HOS estimator to the data-aided benchmark under timing offset, carrier frequency offset (CFO), and phase offset. To avoid ambiguity between common synchronization errors and multiple comparable propagation components, we consider a single dominant path having the ground-truth parameters of $\tau=30,\mathrm{ns}$, $v=12,\mathrm{m/s}$, and $\theta=8^\circ$. The impairments are injected directly into the received tensor: a timing offset $\tau_{\mathrm{sync}}$ as a subcarrier-dependent phase ramp $e^{-j2\pi \Delta f \tau_{\mathrm{sync}} k}$, a CFO $\nu_{\mathrm{sync}}$ as a symbol-dependent phase rotation $e^{j2\pi T_{\mathrm{sym}}\nu_{\mathrm{sync}} n}$, and a constant phase offset as a common rotation $e^{j\phi_0}$. Unless otherwise stated, the SNR is fixed at $10,\mathrm{dB}$, with sweep ranges of $\tau_{\mathrm{sync}}\in\{0,1,2,5,10,20,40\}\ \mathrm{ns}$, $\nu_{\mathrm{sync}}\in\{0,20,50,100,200,400,800\}\ \mathrm{Hz}$, and $\phi_0\in\{0,10,20,40,80,120,180\}^\circ$. Fig.~\ref{fig:sync_timing} shows that the delay RMSE increases with $\tau_{\mathrm{sync}}$, since a timing offset induces the same subcarrier-dependent phase structure as the physical propagation delay and therefore appears as an effective delay shift. By contrast, the velocity and angle RMSE remain nearly unchanged because $\tau_{\mathrm{sync}}$ does not affect the phase evolution across OFDM symbols or antennas. Fig.~\ref{fig:sync_cfo} shows that the velocity RMSE increases with $\nu_{\mathrm{sync}}$, since the CFO produces a symbol-to-symbol phase rotation indistinguishable from Doppler in the model adopted, whereas the delay and angle estimates remain comparatively stable. Fig.~\ref{fig:sync_phase} further shows that both estimators are rather insensitive to a constant phase offset, because the estimation relies mainly on spectral peak locations and matched-inner-product magnitudes, which are invariant to a global phase rotation. Overall, the proposed blind HOS method exhibits the same qualitative sensitivity to synchronization impairments as the data-aided benchmark: degradation occurs primarily when the impairment introduces a phase structure identical to that of a physical parameter, resulting in an effective parameter shift, rather than a general loss of estimation efficiency.



\subsection{Experimental Results}

In the experiment, we employ the SDR-based ISAC platform shown in Fig.~\ref{fig:experiment_setting}. 
The testbed comprises separated transmit and receive nodes. 
At the transmitter, a single SDR provides two RF chains: one chain radiates the sensing OFDM waveform, while the other generates a co-channel interfering OFDM waveform through an independent RF front-end. 
To emulate distinct propagation conditions at the receiver, we introduce a controllable baseband delay and Doppler (velocity) offset between the two RF chains, corresponding to a pair of effective sources: (i) a target at \(3\,\mathrm{m}\) with \(15\,\mathrm{m/s}\) radial velocity and (ii) an interfering source at \(12\,\mathrm{m}\) with \(-20\,\mathrm{m/s}\) radial velocity. 
The receive SDR, placed at the opposite side of the room, records the composite baseband signal for offline processing and parameter estimation. 
Unless otherwise specified, the receiver gain is fixed to \(30\,\mathrm{dB}\) and the transmitter gain is set to \(20\,\mathrm{dB}\).

Fig.~\ref{fig:exp_range_velocity_heatmap} presents the measured range--velocity map obtained from the proposed blind HOS-based estimator. 
The two dominant peaks align well with the programmed delay/velocity settings, indicating that the method can accurately recover both the target and the co-channel interferer parameters from the composite measurement. 
Compared to the simulation results, the experimental map exhibits elevated sidelobes and a higher background level, which we attribute to practical imperfections such as residual carrier-frequency offsets and multipath reflections in the indoor environment. 
Despite these impairments, the proposed approach remains effective, demonstrating its robustness under realistic hardware and propagation conditions.

\section{Conclusions}
\label{sec:conclusion}

Although interference can severely corrupt OFDM-based communication on the time-frequency grid, the underlying sensing parameters remain recoverable through their structured signatures across frequency, slow time, and space. Based on this observation, we developed a fully blind receiver framework using fourth-order tensor construction and a 3D HOS periodogram for joint delay, Doppler, and angle estimation under strong unknown interference. Most importantly, we showed that asymmetric constellation design is the key enabler for blind demodulation, since it resolves the residual phase-rotation and permutation ambiguities left by blind sensing and separation, thereby making coherent demodulation possible without pilots or reference channels. Numerical and experimental results further showed that the constellation design directly controls the sensing-communication tradeoff, with the balanced asymmetric design achieving the best overall end-to-end performance. These results establish asymmetric-constellation-aided HOS processing as an effective framework for robust OFDM-ISAC in interference-intensive environments.

\appendices
\section{Derivation of the Cramér-Rao Lower Bounds}
\label{app:crlb_derivation}

\subsection{Signal Model and Fisher Information Matrix}
Consider the received signal model for a single source (target) after vectorization:
\begin{equation}
    \mathbf{y} = \alpha \mathbf{s}(\bm{\xi}) \odot \mathbf{x} + \mathbf{w},
\end{equation}
where $\mathbf{y} \in \mathbb{C}^{L \times 1}$ with $L = K N_{sym} M$. $\odot$ denotes the element-wise product. The geometric parameter vector is $\bm{\xi} = [\tau, \nu, \theta]^T$, and $\alpha$ is the complex amplitude. The vector $\mathbf{x}$ contains the known transmitted symbols, assumed to have constant modulus $|\mathbf{x}_i|^2 = \sigma_x^2$. The noise $\mathbf{w}$ is circular complex Gaussian with variance $\sigma_n^2$.

The Fisher Information Matrix (FIM) for the full parameter set $\bm{\theta} = [\tau, \nu, \theta, \Re\{\alpha\}, \Im\{\alpha\}]^T$ is block-diagonal under the assumption of a centered observation grid. Specifically, the cross-terms between geometric parameters and amplitude vanish, because the steering vector $\mathbf{s}$ and its derivative $\frac{\partial \mathbf{s}}{\partial \xi}$ are orthogonal (i.e., $\mathbf{s}^H \frac{\partial \mathbf{s}}{\partial \xi} = 0$). Thus, we can evaluate the bounds for $\bm{\xi}$ and $\alpha$ independently.

The diagonal elements of the FIM for a parameter $\psi$ are given by:
\begin{equation}
    J_{\psi\psi} = \frac{2}{\sigma_n^2} \left\| \frac{\partial \mathbf{\mu}}{\partial \psi} \right\|^2, \quad \text{where } \mathbf{\mu} = \alpha \mathbf{s}(\bm{\xi}) \odot \mathbf{x}.
\end{equation}

\subsection{Fisher Information for Delay ($\tau$)}
The delay parameter $\tau$ appears only in the frequency steering vector $\mathbf{a}_f$. Upon using centered indices $k \in [-\frac{K-1}{2}, \dots, \frac{K-1}{2}]$, the derivative becomes:
\begin{equation}
    \frac{\partial [\mathbf{a}_f]_k}{\partial \tau} = (-j 2\pi \Delta f k) \cdot e^{-j 2\pi \Delta f k \tau}.
\end{equation}
The squared norm of the signal derivative is:
\begin{align}
    \left\| \frac{\partial \mathbf{\mu}}{\partial \tau} \right\|^2 &= |\alpha|^2 \sigma_x^2 \cdot (M N_{sym}) \cdot \sum_{k} \left| -j 2\pi \Delta f k \right|^2 \nonumber \\
    &= |\alpha|^2 \sigma_x^2 G_{tot} (2\pi \Delta f)^2 \frac{K^2-1}{12}.
\end{align}
Substituting this into the FIM and using $\text{SNR} = \frac{|\alpha|^2 \sigma_x^2}{\sigma_n^2}$ yields:
\begin{equation}
    \text{CRLB}(\tau) = \frac{1}{J_{\tau\tau}} = \frac{1}{2 \cdot \text{SNR} \cdot G_{tot} \cdot \beta_f^2},
\end{equation}
where $\beta_f^2 = \frac{(2\pi \Delta f)^2 (K^2-1)}{12}$ is the effective bandwidth squared.

\subsection{Fisher Information for Velocity ($\nu$)}
The Doppler velocity $\nu$ appears in the temporal steering vector $\mathbf{a}_t$. The derivative w.r.t. $\nu$ (using $f_D = \nu/\lambda$) is:
\begin{equation}
    \frac{\partial [\mathbf{a}_t]_n}{\partial \nu} = \left( j 2\pi T_{sym} n \frac{1}{\lambda} \right) \cdot e^{j 2\pi T_{sym} n \frac{\nu}{\lambda}}.
\end{equation}
The squared norm of the signal derivative is:
\begin{equation}
    \left\| \frac{\partial \mathbf{\mu}}{\partial \nu} \right\|^2 = |\alpha|^2 \sigma_x^2 G_{tot} \left( \frac{2\pi T_{sym}}{\lambda} \right)^2 \frac{N_{sym}^2-1}{12}.
\end{equation}
The CRLB for velocity is formulated:
\begin{equation}
    \text{CRLB}(\nu) = \frac{1}{2 \cdot \text{SNR} \cdot G_{tot} \cdot \beta_t^2} \cdot \lambda^2,
\end{equation}
where $\beta_t^2 = \frac{(2\pi T_{sym})^2 (N_{sym}^2-1)}{12}$ is the effective time duration squared.

\subsection{Fisher Information for Angle ($\theta$)}
The angle $\theta$ appears in the spatial steering vector $\mathbf{a}_s$. The derivative is:
\begin{equation}
    \frac{\partial [\mathbf{a}_s]_m}{\partial \theta} = \left( -j 2\pi \frac{d}{\lambda} m \cos(\theta) \right) \cdot e^{-j 2\pi \frac{d}{\lambda} m \sin(\theta)}.
\end{equation}
The squared norm of the signal derivative is:
\begin{equation}
    \left\| \frac{\partial \mathbf{\mu}}{\partial \theta} \right\|^2 = |\alpha|^2 \sigma_x^2 G_{tot} \left( 2\pi \frac{d}{\lambda} \cos \theta \right)^2 \frac{M^2-1}{12}.
\end{equation}
The CRLB for the angle (in radians) is expressed as:
\begin{equation}
    \text{CRLB}(\theta) = \frac{1}{2 \cdot \text{SNR} \cdot G_{tot} \cdot \beta_s^2 \cdot \cos^2(\theta)},
\end{equation}
where $\beta_s^2 = \frac{(2\pi d/\lambda)^2 (M^2-1)}{12}$.

\subsection{Stochastic (Blind) CRLB Behavior}
\label{app:stochastic_crb}


For a blind receiver with finite-alphabet data, the exact likelihood is non-Gaussian. A common tractable surrogate is to model the unknown symbols as i.i.d.\ circular Gaussian,
$\mathbf{x}\sim\mathcal{CN}(\mathbf{0},\sigma_x^2\mathbf{I})$.
Then the stacked observation is also complex Gaussian,
$\mathbf{y}\sim\mathcal{CN}(\mathbf{0},\mathbf{R})$, with covariance of:
\begin{equation}
\mathbf{R}(\boldsymbol{\xi})=\sigma_x^2\,\mathbf{s}(\boldsymbol{\xi})\mathbf{s}(\boldsymbol{\xi})^H+\sigma_n^2\mathbf{I},
\end{equation}
where $\mathbf{s}(\boldsymbol{\xi})\in\mathbb{C}^{L}$ denotes the (deterministic) sensing steering vector parameterized by $\boldsymbol{\xi}$ and $L$ is the total number of space--time--frequency samples (e.g., $L=K\,N_{\rm sym}\,M$).

For circular complex Gaussian observations, the Slepian--Bangs formula~\cite{kay1993fundamentals} gives the stochastic FIM as
\begin{equation}
J^{(\mathrm{sto})}_{ij}
=\mathrm{Tr}\!\left(\mathbf{R}^{-1}\mathbf{R}_{i}\mathbf{R}^{-1}\mathbf{R}_{j}\right),
\qquad
\mathbf{R}_{i}\triangleq\frac{\partial \mathbf{R}}{\partial \xi_i}.
\end{equation}
Let $\mathbf{s}_i\triangleq \partial \mathbf{s}/\partial \xi_i$ and define the per-snapshot SNR as
$\mathrm{SNR}\triangleq \sigma_x^2/\sigma_n^2$ and the effective SNR as
\begin{equation}
\mathrm{SNR}_{\mathrm{eff}}\triangleq \frac{\sigma_x^2}{\sigma_n^2}\,\|\mathbf{s}\|^2
= \mathrm{SNR}\cdot \|\mathbf{s}\|^2.
\end{equation}
Upon using the matrix inversion lemma, we have:
\begin{equation}
\mathbf{R}^{-1}
=\frac{1}{\sigma_n^2}\left(
\mathbf{I}
-\frac{\sigma_x^2}{\sigma_n^2+\sigma_x^2\|\mathbf{s}\|^2}\,\mathbf{s}\mathbf{s}^H
\right).
\label{eq:Rinv_woodbury}
\end{equation}
Moreover,
\begin{equation}
\mathbf{R}_i=\sigma_x^2\left(\mathbf{s}_i\mathbf{s}^H+\mathbf{s}\mathbf{s}_i^H\right).
\end{equation}

For a single source, after substituting \eqref{eq:Rinv_woodbury} and simplifying, the stochastic FIM for the sensing parameters $\boldsymbol{\xi}$ admits the compact scaling
\begin{equation}
J^{(\mathrm{sto})}_{ij}
=
\frac{\mathrm{SNR}_{\mathrm{eff}}}{1+\mathrm{SNR}_{\mathrm{eff}}}\;
J^{(\mathrm{DA})}_{ij},
\label{eq:sto_vs_da_scaling}
\end{equation}
where $J^{(\mathrm{DA})}_{ij}$ denotes the corresponding data-aided FIM (i.e., the benchmark when the waveform is effectively known). Equation~\eqref{eq:sto_vs_da_scaling} highlights that the blind (stochastic) information is reduced by the factor $\mathrm{SNR}_{\mathrm{eff}}/(1+\mathrm{SNR}_{\mathrm{eff}})$. At high effective SNR, $\mathrm{SNR}_{\mathrm{eff}}\gg 1$, we have
\begin{equation}
\frac{\mathrm{SNR}_{\mathrm{eff}}}{1+\mathrm{SNR}_{\mathrm{eff}}}
=1-\frac{1}{1+\mathrm{SNR}_{\mathrm{eff}}}
\approx 1-\frac{1}{\mathrm{SNR}_{\mathrm{eff}}}.
\end{equation}
Since $\|\mathbf{s}\|^2$ typically scales linearly with the total dimension $L$ (e.g., $\|\mathbf{s}\|^2\approx L$ under unit-modulus steering), one has $\mathrm{SNR}_{\mathrm{eff}}\approx L\cdot \mathrm{SNR}$, so the correction term becomes negligible in large-scale ISAC integrations. Consequently, under the Gaussian-symbol surrogate, the stochastic CRLB approaches the data-aided CRLB as $L\cdot \mathrm{SNR}$ becomes large.

\bibliographystyle{IEEEtran}
\bibliography{main}

@ARTICLE{ISAC_applications,
  author={Cui, Yuanhao and Liu, Fan and Jing, Xiaojun and Mu, Junsheng},
  journal={IEEE Network}, 
  title={Integrating Sensing and Communications for Ubiquitous {IoT}: Applications, Trends, and Challenges}, 
  year={2021},
  volume={35},
  number={5},
  pages={158-167},
  doi={10.1109/MNET.010.2100152}}

@ARTICLE{ISAC_interfrence,
  author={Niu, Yangyang and Wei, Zhiqing and Wang, Lin and Wu, Huici and Feng, Zhiyong},
  journal={IEEE Internet of Things Journal}, 
  title={Interference Management for Integrated Sensing and Communication Systems: A Survey}, 
  year={2025},
  volume={12},
  number={7},
  pages={8110-8134},
  doi={10.1109/JIOT.2024.3506162}}

@ARTICLE{ISAC_SP,
  author={Zhang, J. Andrew and Liu, Fan and Masouros, Christos and Heath, Robert W. and Feng, Zhiyong and Zheng, Le and Petropulu, Athina},
  journal={IEEE Journal of Selected Topics in Signal Processing}, 
  title={An Overview of Signal Processing Techniques for Joint Communication and Radar Sensing}, 
  year={2021},
  volume={15},
  number={6},
  pages={1295-1315},
  doi={10.1109/JSTSP.2021.3113120}}

@article{ZHANG2023103861,
title = {Joint beampattern and symbol-level waveform design for integrated sensing and communication},
journal = {Digital Signal Processing},
volume = {133},
pages = {103861},
year = {2023},
issn = {1051-2004},
doi = {https://doi.org/10.1016/j.dsp.2022.103861},
author = {Zhibo Zhang and Qing Chang and Leyan Chen and Zichu Zhao}
}

@ARTICLE{Wang2025_SLP_ISAC,
  author={Wang, Yiran and Hu, Xiaoyan and Li, Ang and Masouros, Christos and Wong, Kai-Kit and Yang, Kun},
  journal={IEEE Transactions on Wireless Communications}, 
  title={Symbol-Scaling Based Interference Exploitation in {ISAC} Systems: From Symbol Level to Block Level}, 
  year={2025},
  volume={24},
  number={3},
  pages={2451-2466},
  doi={10.1109/TWC.2024.3521584}}

@ARTICLE{Chen2023_SLP_MIMO_ISAC,
  author={Chen, Yunwang and Liu, Fan and Liao, Zihan and Dong, Fuwang},
  journal={IEEE Communications Letters}, 
  title={Symbol-Level Precoding for {MIMO} {ISAC} Transmission Based on Interference Exploitation}, 
  year={2024},
  volume={28},
  number={2},
  pages={283-287},
  doi={10.1109/LCOMM.2023.3346884}}

@ARTICLE{Liao2023_SLP_ISAC,
  author={Liao, Zihan and Liu, Fan},
  journal={IEEE Communications Letters}, 
  title={Symbol-Level Precoding for Integrated Sensing and Communications: A Faster-Than-Nyquist Approach}, 
  year={2023},
  volume={27},
  number={12},
  pages={3210-3214},
  doi={10.1109/LCOMM.2023.3328140}}

@INPROCEEDINGS{Cai2024_SLP_SIC_ISAC,
  author={Cai, Shu and Chen, Zihao and Liu, Ya-Feng and Zhang, Jun},
  booktitle={IEEE International Conference on Acoustics, Speech and Signal Processing (ICASSP)}, 
  title={Symbol-Level Precoding-Based Self-Interference Cancellation for {ISAC} Systems}, 
  year={2025},
  volume={},
  number={},
  pages={1-5},
  doi={10.1109/ICASSP49660.2025.10887946}}

@ARTICLE{Wei2025_OTFS_V2X_ISAC,
  author={Wei, Xinyuan and Yuan, Weijie and Zhang, Kecheng and Liu, Fan},
  journal={IEEE Transactions on Mobile Computing}, 
  title={{OTFS}-Assisted {ISAC} System: Delay Doppler Channel Estimation and SDR-Based Implementation}, 
  year={2025},
  volume={24},
  number={11},
  pages={11865-11878},
  doi={10.1109/TMC.2025.3582421}}

@article{Shi2024_OTFS_ISAC,
title = {Reliability performance analysis for {OTFS} modulation based integrated sensing and communication},
journal = {Digital Signal Processing},
volume = {144},
pages = {104280},
year = {2024},
issn = {1051-2004},
doi = {https://doi.org/10.1016/j.dsp.2023.104280},
author = {Jia Shi and Xinyang Hu and Zhuangzhuang Tie and Xuankai Chen and Wei Liang and Zan Li}
}

@article{sun2014cyclostationarity,
  title={Cyclostationarity-based joint domain approach to blind recognition of {SCLD} and {OFDM} signals},
  author={Sun, Zhuo and Chen, Yilong and Liu, Siyuan and Wang, Wenbo},
  journal={EURASIP Journal on Advances in Signal Processing},
  volume={2014},
  number={1},
  pages={5},
  year={2014},
  publisher={Springer}
}

@ARTICLE{DBD,
  author={Vargas, Edwin and Mishra, Kumar Vijay and Jacome, Roman and Sadler, Brian M. and Arguello, Henry},
  journal={IEEE Journal on Selected Areas in Information Theory}, 
  title={Dual-Blind Deconvolution for Overlaid Radar-Communications Systems}, 
  year={2023},
  volume={4},
  number={},
  pages={75-93},
  doi={10.1109/JSAIT.2023.3287823}}

@article{Jacome2024_MDBD,
title = {Multi-antenna dual-blind deconvolution for joint radar-communications via SoMAN minimization},
journal = {Signal Processing},
volume = {221},
pages = {109484},
year = {2024},
issn = {0165-1684},
doi = {https://doi.org/10.1016/j.sigpro.2024.109484},
author = {Roman Jacome and Edwin Vargas and Kumar Vijay Mishra and Brian M. Sadler and Henry Arguello}
}

@INPROCEEDINGS{subspace_problem1,
  author={Badeau, R. and David, B. and Richard, G.},
  booktitle={IEEE International Conference on Acoustics, Speech, and Signal Processing}, 
  title={Selecting the modeling order for the {ESPRIT} high resolution method: an alternative approach}, 
  year={2004},
  volume={2},
  number={},
  pages={ii-1025},
  doi={10.1109/ICASSP.2004.1326435}}

@INPROCEEDINGS{subspace_problem2,
  author={Salman, Tara and Badawy, Ahmed and Elfouly, Tarek M. and Mohamed, Amr and Khattab, Tamer},
  booktitle={International Wireless Communications and Mobile Computing Conference (IWCMC)}, 
  title={Estimating the number of sources: An efficient maximization approach}, 
  year={2015},
  volume={},
  number={},
  pages={199-204},
  doi={10.1109/IWCMC.2015.7289082}}

@book{kay1993fundamentals,
  title={Fundamentals of statistical signal processing: estimation theory},
  author={Kay, Steven M},
  year={1993},
  publisher={Prentice-Hall, Inc.}
}

@ARTICLE{
  ISAC1,
  author={Liu, Fan and Cui, Yuanhao and Masouros, Christos and Xu, Jie and Han, Tony Xiao and Eldar, Yonina C. and Buzzi, Stefano},
  journal={IEEE Journal on Selected Areas in Communications}, 
  title={Integrated Sensing and Communications: Toward Dual-Functional Wireless Networks for 6{G} and Beyond}, 
  year={2022},
month={Mar.},
  volume={40},
  number={6},
  pages={1728-1767},
  doi={10.1109/JSAC.2022.3156632}}

@ARTICLE{ISAC2,
  author={Lu, Shihang and Liu, Fan and Li, Yunxin and Zhang, Kecheng and Huang, Hongjia and Zou, Jiaqi and Li, Xinyu and Dong, Yuxiang and Dong, Fuwang and Zhu, Jia and Xiong, Yifeng and Yuan, Weijie and Cui, Yuanhao and Hanzo, Lajos},
  journal={IEEE Internet of Things Journal}, 
  title={Integrated Sensing and Communications: Recent Advances and Ten Open Challenges}, 
  year={2024},
month={Feb.},
  volume={11},
  number={11},
  pages={19094-19120},
  doi={10.1109/JIOT.2024.3361173}}

@ARTICLE{11479627,
  author={Pu, Henglin and Wang, Xuefeng and Su, Lu and Li, Husheng},
  journal={IEEE Transactions on Wireless Communications}, 
  title={Space-Time-Frequency Synthetic Integrated Sensing and Communication Networks}, 
  year={2026},
  volume={25},
  number={},
  pages={15004-15020},
  doi={10.1109/TWC.2026.3681132}}

@ARTICLE{11432914,
  author={Pu, Henglin and Han, Zhu and Petropulu, Athina P. and Li, Husheng},
  journal={IEEE Wireless Communications Letters}, 
  title={Constellation-Based Blind Sensing for {OFDM}-{ISAC} With Cochannel Interference}, 
  year={2026},
  volume={15},
  number={},
  pages={2338-2342},
  doi={10.1109/LWC.2026.3673175}}

@ARTICLE{NOMA,
  author={Ding, Zhiguo and Schober, Robert and Poor, H. Vincent},
  journal={IEEE Transactions on Wireless Communications}, 
  title={A General {MIMO} Framework for {NOMA} Downlink and Uplink Transmission Based on Signal Alignment}, 
  year={2016},
  volume={15},
  number={6},
  pages={4438-4454},
  doi={10.1109/TWC.2016.2542066}}

@INPROCEEDINGS{9746868,
  author={Vargas, Edwin and Mishra, Kumar Vijay and Jacome, Roman and Sadler, Brian M. and Arguello, Henry},
  booktitle={IEEE International Conference on Acoustics, Speech and Signal Processing (ICASSP)}, 
  title={Joint Radar-Communications Processing from A Dual-Blind Deconvolution Perspective}, 
  year={2022},
  volume={},
  number={},
  pages={5622-5626},
  doi={10.1109/ICASSP43922.2022.9746868}}

@article{pu2025wideband,
  title={Wideband Integrated Sensing and Communications: Spectral Efficiency and Signaling Design},
  author={Pu, Henglin and Han, Zhu and Petropulu, Athina P and Li, Husheng},
  journal={arXiv preprint arXiv:2509.24097},
  year={2025}
}

@ARTICLE{1092165,
  author={Thomas, C. and Weidner, M. and Durrani, S.},
  journal={IEEE Transactions on Communications}, 
  title={Digital Amplitude-Phase Keying with {M}-Ary Alphabets}, 
  year={1974},
  volume={22},
  number={2},
  pages={168-180},
  doi={10.1109/TCOM.1974.1092165}}

@ARTICLE{bolcskei2002blind,
  author={Bolcskei, H. and Heath, R.W. and Paulraj, A.J.},
  journal={IEEE Transactions on Signal Processing}, 
  title={Blind channel identification and equalization in {OFDM}-based multiantenna systems}, 
  year={2002},
  month   = {jan.},
  volume={50},
  number={1},
  pages={96-109},
  doi={10.1109/78.972486}}

@ARTICLE{muquet2002subspace,
  author={Muquet, B. and de Courville, M. and Duhamel, P.},
  journal={IEEE Transactions on Signal Processing}, 
  title={Subspace-based blind and semi-blind channel estimation for {OFDM} systems}, 
  year={2002},
  month   = {jul.},
  volume={50},
  number={7},
  pages={1699-1712},
  doi={10.1109/TSP.2002.1011210}}

@ARTICLE{yuan2024integrated,
  author={Yuan, Zhengdao and Guo, Qinghua and Eldar, Yonina C. and Li, Yonghui},
  journal={IEEE Transactions on Communications}, 
  title={Integrated Near Field Sensing and Communications Using Unitary Approximate Message Passing-Based Matrix Factorization}, 
  year={2025},
  month={Oct.},
  volume={73},
  number={10},
  pages={9939-9953},
  doi={10.1109/TCOMM.2025.3582730}}

@ARTICLE{Gupta2020_BlindOFDMModClass,
  author={Gupta, Rahul and Kumar, Sushant and Majhi, Sudhan},
  journal={IEEE Transactions on Vehicular Technology}, 
  title={Blind Modulation Classification for Asynchronous {OFDM} Systems Over Unknown Signal Parameters and Channel Statistics}, 
  year={2020},
  month={May},
  volume={69},
  number={5},
  pages={5281-5292},
  doi={10.1109/TVT.2020.2981935}}

@ARTICLE{Valiulahi2026_LANM_ISAC,
  author={Valiulahi, Iman and Masouros, Christos and Petropulu, Athina P.},
  journal={IEEE Transactions on Communications}, 
  title={{ISAC} Super-Resolution Receiver via Lifted Atomic Norm Minimization}, 
  year={2026},
  volume={74},
  number={},
  pages={5184-5198},
  doi={10.1109/TCOMM.2026.3663525}}

@ARTICLE{1455106,
  author={Harris, F.J.},
  journal={Proceedings of the IEEE}, 
  title={On the use of windows for harmonic analysis with the discrete Fourier transform}, 
  year={1978},
  month={Jan.},
  volume={66},
  number={1},
  pages={51-83},
  doi={10.1109/PROC.1978.10837}}

@ARTICLE{4102829,
  author={Rohling, Hermann},
  journal={IEEE Transactions on Aerospace and Electronic Systems}, 
  title={Radar {CFAR} Thresholding in Clutter and Multiple Target Situations}, 
  year={1983},
  month = {Jul.},
  volume={AES-19},
  number={4},
  pages={608-621},
  doi={10.1109/TAES.1983.309350}}

\end{document}